\documentclass[iop,revtex4]{emulateapj}
\slugcomment{Accepted to ApJ: October 28, 2015}

\usepackage[backref,breaklinks,colorlinks,citecolor=blue]{hyperref}
\usepackage[all]{hypcap}
\usepackage{graphicx} 
\usepackage{mathrsfs} 
\usepackage{natbib}
\usepackage{amsmath}
\usepackage{float}
\usepackage{amssymb}
\usepackage{color}

\usepackage[hang,multiple]{footmisc}

\newcommand{\oi}{[O\,{\sc i}]}
\newcommand{\ha}{H$\alpha$}

\newcommand{\as}{$^{\prime\prime}$}

\begin{document}
\title{Young ``Dipper" Stars in Upper Sco and $\rho$ Oph Observed by K2}

\author{M. Ansdell\altaffilmark{1,10}, E. Gaidos\altaffilmark{2,3,4,10}, S. A. Rappaport\altaffilmark{5}, T. L. Jacobs\altaffilmark{6}, D. M. LaCourse\altaffilmark{6}, K. J. Jek\altaffilmark{6}, A. W. Mann\altaffilmark{7,10}, M. C. Wyatt\altaffilmark{8}, G. Kennedy\altaffilmark{8}, J. P. Williams\altaffilmark{1}, T. S. Boyajian\altaffilmark{9}}

\altaffiltext{1}{Institute for Astronomy, University of Hawai`i at M\={a}noa, Honolulu, HI}
\altaffiltext{2}{Department of Geology \& Geophysics, University of Hawai`i at M\={a}noa, Honolulu, HI}
\altaffiltext{3}{Visiting Scientist, Sauverny Observatoire,  l'Universit\'{e} de Gen\`{e}ve}
\altaffiltext{4}{Visiting Scientist, Center for Astrophysics, Harvard University, Cambridge, MA}
\altaffiltext{5}{Physics Department and Kavli Institute for Astrophysics and Space Research, Massachusetts Institute of Technology, Cambridge, MA}
\altaffiltext{6}{Amateur Astronomer}
\altaffiltext{7}{Harlan J. Smith Fellow, Department of Astronomy, The University of Texas at Austin, Austin, TX}
\altaffiltext{8}{Institute of Astronomy, University of Cambridge, Madingley Road, Cambridge, UK}
\altaffiltext{9}{Department of Astronomy, Yale University, New Haven, CT}
\altaffiltext{10}{Visiting Astronomer, NASA Infrared Telescope Facility, operated by the University of Hawaii under contract NNH14CK55B}


\begin{abstract}
We present ten young ($\lesssim$10 Myr) late-K and M dwarf stars observed in K2 Campaign 2 that host protoplanetary disks and exhibit quasi-periodic or aperiodic dimming events. Their optical light curves show $\sim$10--20 dips in flux over the 80-day observing campaign with durations of $\sim$0.5--2 days and depths of up to $\sim$40\%. These stars are all members of the $\rho$ Ophiuchus ($\sim$1 Myr) or Upper Scorpius ($\sim$10 Myr) star-forming regions. To investigate the nature of these ``dippers" we obtained: optical and near-infrared spectra to determine stellar properties and identify accretion signatures; adaptive optics imaging to search for close companions that could cause optical variations and/or influence disk evolution; and millimeter-wavelength observations to constrain disk dust and gas masses. The spectra reveal Li {\sc I} absorption and H$\alpha$ emission consistent with stellar youth (\textless50 Myr), but also accretion rates spanning those of classical and weak-line T Tauri stars. Infrared excesses are consistent with protoplanetary disks extending to within $\sim$10 stellar radii in most cases; however, the sub-mm observations imply disk masses that are an order of magnitude below those of typical protoplanetary disks. We find a positive correlation between dip depth and WISE-2 excess, which we interpret as evidence that the dipper phenomenon is related to occulting structures in the inner disk, although this is difficult to reconcile with the weakly accreting aperiodic dippers. We consider three mechanisms to explain the dipper phenomenon: inner disk warps near the co-rotation radius related to accretion; vortices at the inner disk edge produced by the Rossby Wave Instability; and clumps of circumstellar material related to planetesimal formation.

\end{abstract}

\maketitle


\section{INTRODUCTION\label{sec-intro}}

 \capstartfalse                                                           
\begin{deluxetable*}{lllllrrrrc}                                              
\tabletypesize{\footnotesize}                                           
\centering                                                               
\tablewidth{500pt}                                                         
\tablecaption{Properties of K2/C2 Dippers \label{tab-properties}}                  
\tablecolumns{10}                                                         
\tablehead{                              
   \colhead{EPIC\textsuperscript{a}}                                      
 & \colhead{2MASS\textsuperscript{b}}                                                          
 & \colhead{RA$_{\rm J2000}$}                                                          
 & \colhead{Dec$_{\rm J2000}$}                                                         
 & \colhead{Mem.\textsuperscript{c}}                                                   
 & \colhead{P$_{\rm rot}$\textsuperscript{d}}                                                           
 & \colhead{N$_{\rm dip}$\textsuperscript{e}}                                                   
 & \colhead{R$_{\rm dip}$\textsuperscript{e}} 
 & \colhead{D$_{\rm dip}$\textsuperscript{e}}   
 & \colhead{Var.\textsuperscript{f}}   
 }                                          
\startdata                                                               
203343161  &  16245587-2627181  &  16:24:55.878  &  -26:27:18.10  &  USc  &  2.24  &  14  & 10.2  &  0.12  &  Q  \\              
203410665  &  16253849-2613540  &  16:25:38.484  &  -26:13:53.98  &  Oph  &  4.24  &  17  &   7.7  &  0.30  &  A    \\
203895983  &  16041893-2430392  &  16:04:18.936  &  -24:30:39.34  &  USc  &  2.44  &  10  &   5.5  &  0.11  &  A    \\
203937317  &  16261706-2420216  &  16:26:17.080  &  -24:20:22.02  &  Oph  &  5.44  &  17  & 11.8  &  0.13  &  Q   \\ 
204137184  &  16020517-2331070  &  16:02:05.180  &  -23:31:06.94  &  USc  &  2.64  &  20  &   9.4  &  0.35  &  Q    \\
204630363  &  16100501-2132318  &  16:10:05.019  &  -21:32:31.89  &  USc  &  6.66  &    8  &   7.8  &  0.10  &  A    \\
204757338  &  16072747-2059442  &  16:07:27.462  &  -20:59:44.22  &  USc  &  2.39  &  12  &   7.1  &  0.08  &  Q   \\ 
204932990  &  16115091-2012098  &  16:11:50.928  &  -20:12:09.89  &  USc  &  2.30  &  11  & 15.3   &  0.26  &  A    \\  
205151387  &  16090075-1908526  &  16:09:00.762  &  -19:08:52.70  &  USc  &  9.55  &  15  &  8.8   &  0.31  &  Q    \\
205519771  &  16071403-1702425  &  16:07:14.018  &  -17:02:42.67  &  USc  &  2.46  &  13  & 11.5  &  0.11  &  A    
\enddata    
\tablenotetext{a}{K2 Ecliptic Plane Input Catalog (EPIC) ID.} 
\tablenotetext{b}{2MASS ID associated with the EPIC ID.} 
\tablenotetext{c}{Cluster membership: USc = Upper Sco, Oph = $\rho$ Oph (Section~\ref{sec-membership}).}
\tablenotetext{d}{Inferred rotation period (days) measured from the K2/C2 light curves (Section~\ref{sec-identification}).}
\tablenotetext{e}{Criteria used to select our dipper sample (Section~\ref{sec-identification}).}
\tablenotetext{f}{Variability type: A = aperiodic, Q = quasi-periodic (Section~\ref{sec-classification})}
\end{deluxetable*}                                                        
 \capstartfalse
 
Young stars exhibit diverse photometric variability attributed to a range of physical mechanisms. The bursting events seen in FU Orionis stars are likely due to episodic accretion during the early stages of stellar evolution \citep{1996ARA&A..34..207H}. The assortment of photometric variability observed in classical T Tauri stars (CTTS), whose light curves are complicated by their full disks and ongoing accretion, has been connected to a plethora of mechanisms, such as obscuring inner disk structures and accretion bursts \cite[e.g.,][]{2014AJ....147...82C}. The sinusoidal light curves exhibited by more evolved weak-line T Tauri stars (WTTS), which no longer show strong accretion signatures, are attributed to magnetic spots on the stellar surface rotating with the star. Thus studying photometric variability in young stars can give insight into the structure and evolution of planet-forming disks as well as the stellar activity affecting planets during their early evolution.

Studies of photometric variability in young stellar clusters have utilized space-based optical and infrared (IR) observatories, in particular the Convection, Rotation and Planetary Transits satellite \cite[CoRoT;][]{2006cosp...36.3749B} and the {\it Spitzer} Infrared Array Camera \cite[IRAC;][]{2004ApJS..154...10F}, which provide more precise photometry and longer observing baselines than ground-based observatories. These studies have revealed a distinct class of photometric variables among CTTS populations, the so-called ``dippers" whose light curves exhibit a relatively flat flux continuum but also contain quasi-periodic or aperiodic drops in flux lasting up to a few days.

In particular, the Orion Nebula Cluster (ONC) was observed by \cite{2011ApJ...733...50M} with {\it Spitzer}/IRAC over 40 days and the NGC 2264 region was observed by \cite{2010A&A...519A..88A} with CoRoT over 23 days and also by \cite{2014AJ....147...82C} with {\it Spitzer}/IRAC and CoRoT simultaneously over 30 days. \cite{2011ApJ...733...50M} found that the dips were shallower in the IR compared to the optical, consistent with extinction by dust; thus they interpreted the dips as arising from dusty inner disk structures passing along our line-of-sight (LOS) to the stellar photosphere. \cite{2010A&A...519A..88A} and \cite{2014AJ....147...82C} found such dippers to be fairly common ($\sim$20--30\%) among the CTTS population in NGC 2264. \cite{2015AJ....149..130S} and \cite{2015A&A...577A..11M} later attributed the dippers in NGC 2264 to occultations of the stellar photosphere by disk structures at or near the co-rotation radius.
 
Detailed follow-up investigations of the dipper populations in young star-forming regions are needed to improve our physical understanding of the dipper phenomenon, which in turn can provide a means of probing disk structure and dynamics during key epochs of planet formation. For example, if the dips are indeed produced by occulting inner disk structures, they could provide a rare opportunity to study the inner disk and its role (if any) in planet formation. If the dips are instead related to material further out in the disk, they could provide information on the size distribution of forming planetesimals in debris disks or transiently formed vortices near the inner edges of transition disks.
 
However, the ONC and NGC 2264 are both young and distant clusters. The ONC has an age of $\sim$2 Myr and a distance of $\sim$400 pc \cite[e.g.,][]{2011A&A...534A..83R}, while NGC 2264 has an age of $\sim$3 Myr and a distance of $\sim$760 pc \cite[e.g.,][]{2008hsf1.book..966D}. Their distances limit detailed follow-up observations of their dipper populations (e.g., resolving $\sim$30 AU scales in NGC 2264 to search for close companions requires $\sim$40 mas resolution), while their similar ages make it desirable to observe other star-forming regions at distinctly different ages (e.g., to study the evolution of the dipper phenomenon).

Fortunately, the K2 mission \citep{2014PASP..126..398H} provides a new opportunity for studying photometric variability in young stellar clusters. Although the original mission of the {\it Kepler} spacecraft \citep{2010Sci...327..977B} focused on solar-age stars at several hundred parsecs, the re-purposed K2 mission is studying fields that include nearby young stellar clusters with ages spanning $\sim$1--600 Myr. In particular, the K2 Campaign 2 (K2/C2) field covered the Upper Scorpius (Upper Sco) and $\rho$ Ophiuchus ($\rho$ Oph) regions, which are relatively nearby at $\sim$140 pc \citep{1999AJ....117..354D} and $\sim$120 pc \citep{2008ApJ...675L..29L}, respectively, and have significantly different ages at $\sim$10 Myr \citep{2012ApJ...746..154P} and $\sim$1 Myr \citep{2007ApJ...671.1800A}, respectively. 

This work focuses on a sample of dippers in the Upper Sco and $\rho$ Oph regions found using the publicly available K2/C2 data. We describe our dipper sample selection in Section 2, then present our follow-up observations in Section 3. We analyze these data in Section 4 and identify observed trends and correlations in Section 5. In Section 6, we discuss possible mechanisms driving the dipper phenomenon. We summarize our findings and discuss areas of future work in Section 7.


\section{DIPPER SAMPLE \label{sec-sample}}

\subsection{K2 Light Curve Extraction \label{sec-extraction}}

The primary mission of the {\it Kepler} spacecraft ended in 2013 May, when the failure of two of its four reaction wheels meant the spacecraft could no longer maintain the high photometric precision required for its original science goals. The re-purposed two-wheeled {\it Kepler} mission, K2, utilizes an ecliptic-observing orientation to mitigate pointing drift due to solar radiation pressure. However, quasi-periodic thruster firings are still needed to correct for residual pointing drift throughout the 80-day observing campaign. This spacecraft motion introduces artificial systematics to the K2 light curves as differing amounts of target flux may be lost/contaminated when applying aperture masks to the K2 Target Pixel Files (TPFs). Although artifact mitigation techniques were previously developed for the {\it Kepler} mission \cite[e.g.,][]{2012PASP..124..963K}, these additional pointing variations require new data reduction solutions optimized for K2 \cite[e.g.,][]{2014PASP..126..948V,2015arXiv150707578H}.

We therefore employed a fixed-mask data reduction pipeline based on the open source {\it Kepler} Guest Observer software package \textsc{PyKE}\footnote{\url{http://keplerscience.arc.nasa.gov/PyKE.shtml}}  \citep{2012ascl.soft08004S}, which was updated in 2014 March to provide additional compatibility with K2-derived TPFs. We obtained K2/C2 data from the Mikulski Archive for Space Telescopes\footnote{\url{https://archive.stsci.edu/}} (MAST) in 2015 March, and utilized the Principal Component Analysis (PCA) technique \textsc{KepPCA}{\footnote{KepPCA release notes: \url{http://keplerscience.arc.nasa.gov/ContributedSoftwareKeppca.shtml}}  \citep{2012AJ....143....4H} to extract light curves for each of the 13,344 stellar long cadence TPFs. \textsc{KepPCA} mitigates artifacts by analyzing pixel samples from a given TPF and segregating observed photometric trends associated with individual pixels from those common to all pixels. We extracted source fluxes using square aperture masks centered at the source locations, where mask dimensions were adjusted as a function of EPIC catalog source magnitude. Our pipeline automatically identifies additional centroids in the mask, but does not subsequently adjust the mask to avoid contamination from nearby sources before extracting photometry.

With any de-trending process comes the risk of altering the underlying astrophysical signals. Thus we compared our corrected data to the K2/C2 \textsc{K2SFF}\footnote{\url{https://archive.stsci.edu/prepds/k2sff/}} light curves released to MAST in 2015 May, which have been motion corrected with the Self Field Flattening (SFF) technique described in \cite{2014PASP..126..948V}. The SFF method improves the pointing precision of K2 aperture photometry by correlating observed variability with spacecraft motion, and has revealed subtle long-term drifts in target centroid positions not accounted for in our \textsc{KepPCA} pipeline. Therefore we used the \textsc{KepPCA} light curves when identifying our dipper sample (Section~\ref{sec-identification}), but used the \textsc{K2SFF} light curves for all subsequent analyses; the exception is EPIC 204137184, whose \textsc{K2SFF} light curve appeared corrupted.

\subsection{Dipper Sample Identification \label{sec-identification}}

\capstartfalse
\begin{figure}
\begin{centering}
\includegraphics[width=8.5cm]{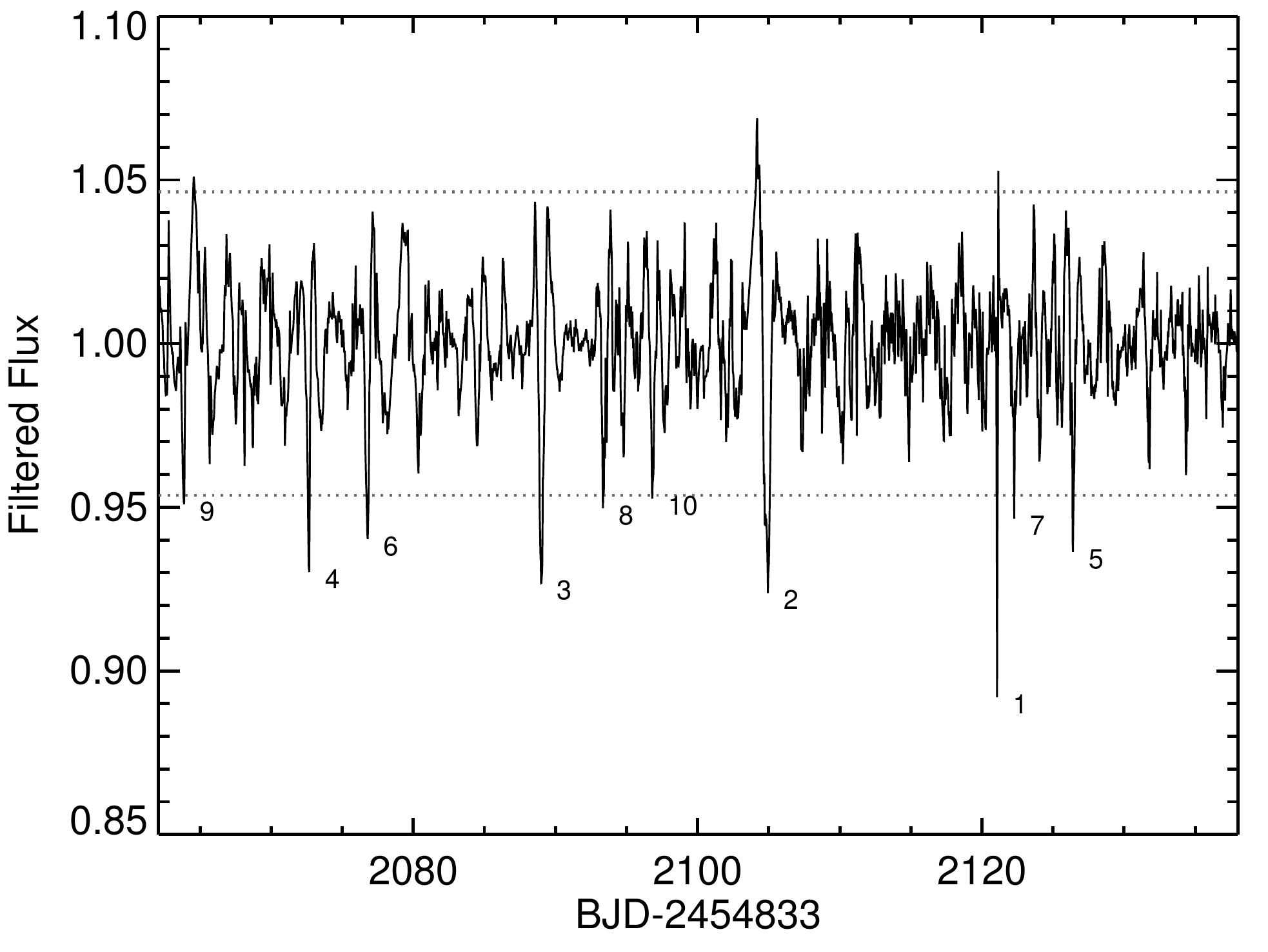}
\caption{\small Filtered light curve for EPIC 203895983. The $x$ axis shows the Barycentric Julian Date (BJD) spanning the 80-day K2/C2 observing period. The $y$ axis shows the normalized flux after applying a high-pass filter (Section~\ref{sec-identification}). The dips in flux below the 3$\sigma$ threshold (bottom dotted line) are numbered in order of dip depth. For comparison, the unfiltered light curve for EPIC 203895983 is shown in Figure~\ref{fig-lcs}.}
\label{fig-hpf}
\end{centering}
\end{figure}
\capstartfalse

We reviewed each of the 13,344 corrected light curves (Section~\ref{sec-extraction}) by eye, flagging $\sim$100 sources exhibiting quasi-periodic or aperiodic dimming events as candidate dippers. The visual inspection was necessary because dippers exhibit complex light curves that make them difficult to identify with automated algorithms. We then conservatively ignored light curves showing suspicious irregularities that appeared to be instrument related (e.g., charge bleed) or the result of data corruption (e.g., data discontinuities) rather than due to any physical phenomenon inherent to the science target. We also ignored particularly noisy light curves that were likely associated with extreme intrinsic stellar variability. For each candidate dipper, we normalized the extracted K2/C2 light curve, then put the normalized light curve through a high-pass filter with a cut-on frequency of 1 day$^{-1}$. This filtering highlighted the quasi-periodic and aperiodic dimming events, while suppressing the periodic variability from stellar rotation, due to their different duty cycles. Figure~\ref{fig-hpf} shows an example of a filtered light curve (to be compared with the normalized, but unfiltered, versions in Figure~\ref{fig-lcs}). We used these light curves to compute several metrics that quantify the dipper phenomenon, and then used the metrics to select our sample of ten dippers presented in Table~\ref{tab-properties}. We describe these metrics and our selection criteria below.

To estimate the inherent variation of each source ($\sigma$) we calculated the Tukey's biweight function \citep{1977eda..book.....T} of the filtered light curve using the {\sc IDL} function \texttt{robust\_sigma}, which gives a weighted standard deviation. The number of dips over the 80-day observing period ($N_{\rm dip}$) was tallied from the drops in flux at least 3$\sigma$ in depth in the filtered light curve (see Figure~\ref{fig-hpf}). We calculated the average depth of the three deepest dips in the normalized light curve ($D_{\rm dip}$), and the ratio of the average depth of the three deepest dips in the filtered light curve to $\sigma$ ($R_{\rm dip}$), as measures of the strength of the dipper phenomenon. We also estimated the stellar rotation period ($P_{\rm rot}$) by computing the autocorrelation function of the normalized light curve, checking our results with Fourier Transform and Lomb-Scargle periodogram techniques to ensure consistency. This assumes that the periodic signal is from stellar rotation and is not influenced by the dimming events; thus although our $P_{\rm rot}$ values are consistent with the stellar rotation periods of young late-type stars \citep[e.g.,][]{2007prpl.conf..297H}, they should be viewed with caution until confirmed by other observations (e.g., $v$ sin $i$ from high-resolution spectra).

To construct our dipper sample, we required N$_{\rm dip}>5$ (to ensure dips were not spurious events), R$_{\rm dip}> 5$ and D$_{\rm dip}> 0.07$ (to identify significant dippers), and P$_{\rm rot}\lesssim10$ days \cite[to confirm stellar youth;][]{2007prpl.conf..297H}. This resulted in a sample of ten dippers, whose properties are listed in Table~\ref{tab-properties} and whose normalized light curves are shown in Figure~\ref{fig-lcs}. We checked for planetary signals using the IPAC Box-Least Squares (BLS) periodogram service\footnote{\url{http://exoplanetarchive.ipac.caltech.edu/applications/Periodogram/}} but did not find any planetary signals.

\capstartfalse
\begin{figure*}
\begin{centering}
\includegraphics[width=18.5cm]{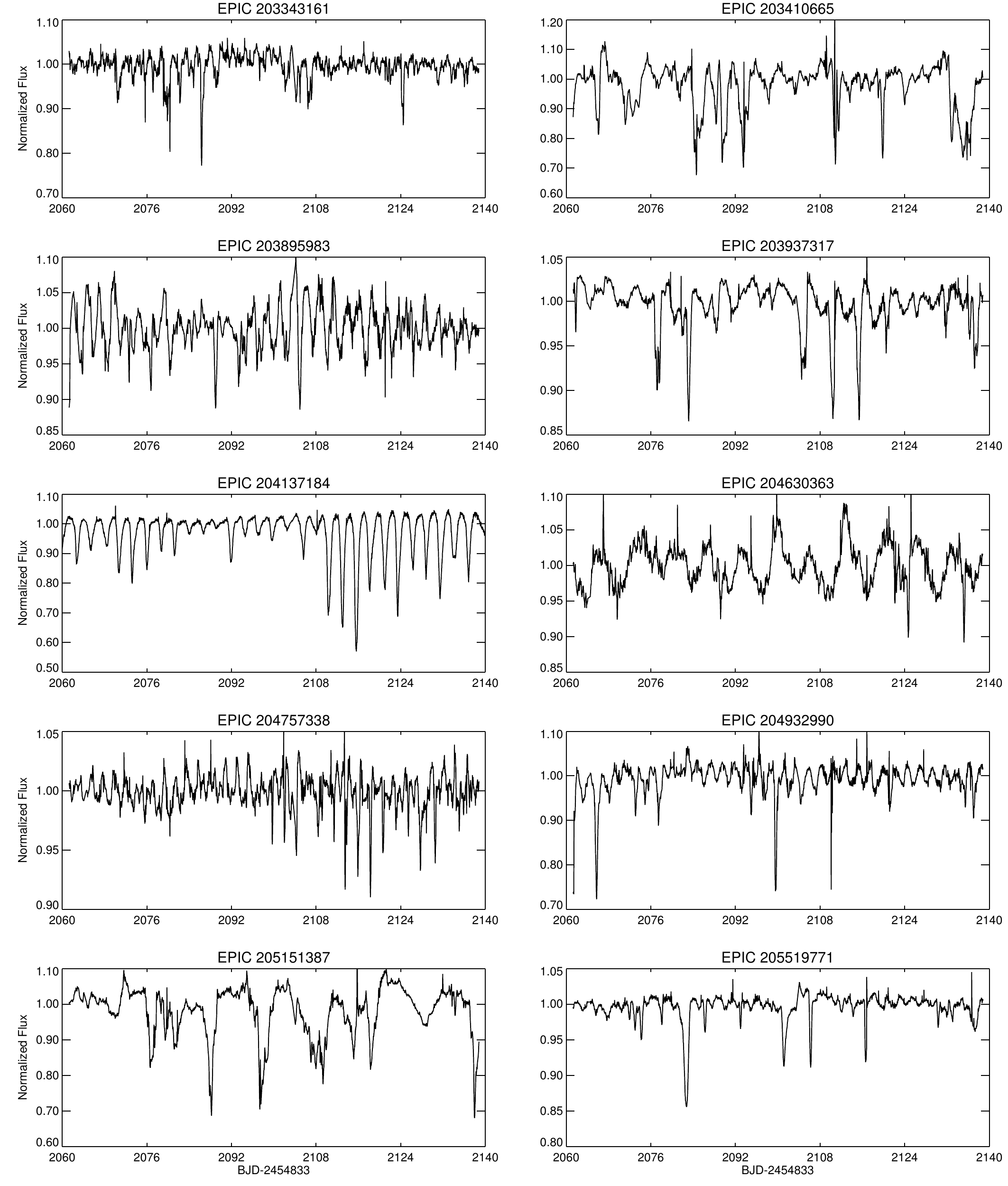}
\caption{\small Normalized light curves (Section~\ref{sec-identification}) for the ten dippers in our sample (Table~\ref{tab-properties}). Photometry is from the publicly available \textsc{K2SFF} light curves, except for EPIC 204137184 whose photometry is from our \textsc{KepPCA} pipeline (Section~\ref{sec-extraction}). The narrow vertical spikes seen in some dippers are likely cosmic rays, data processing artifacts, or stellar flares.}
\label{fig-lcs}
\end{centering}
\end{figure*}
\capstartfalse

\capstartfalse
\begin{figure}
\begin{centering}
\includegraphics[width=8.5cm]{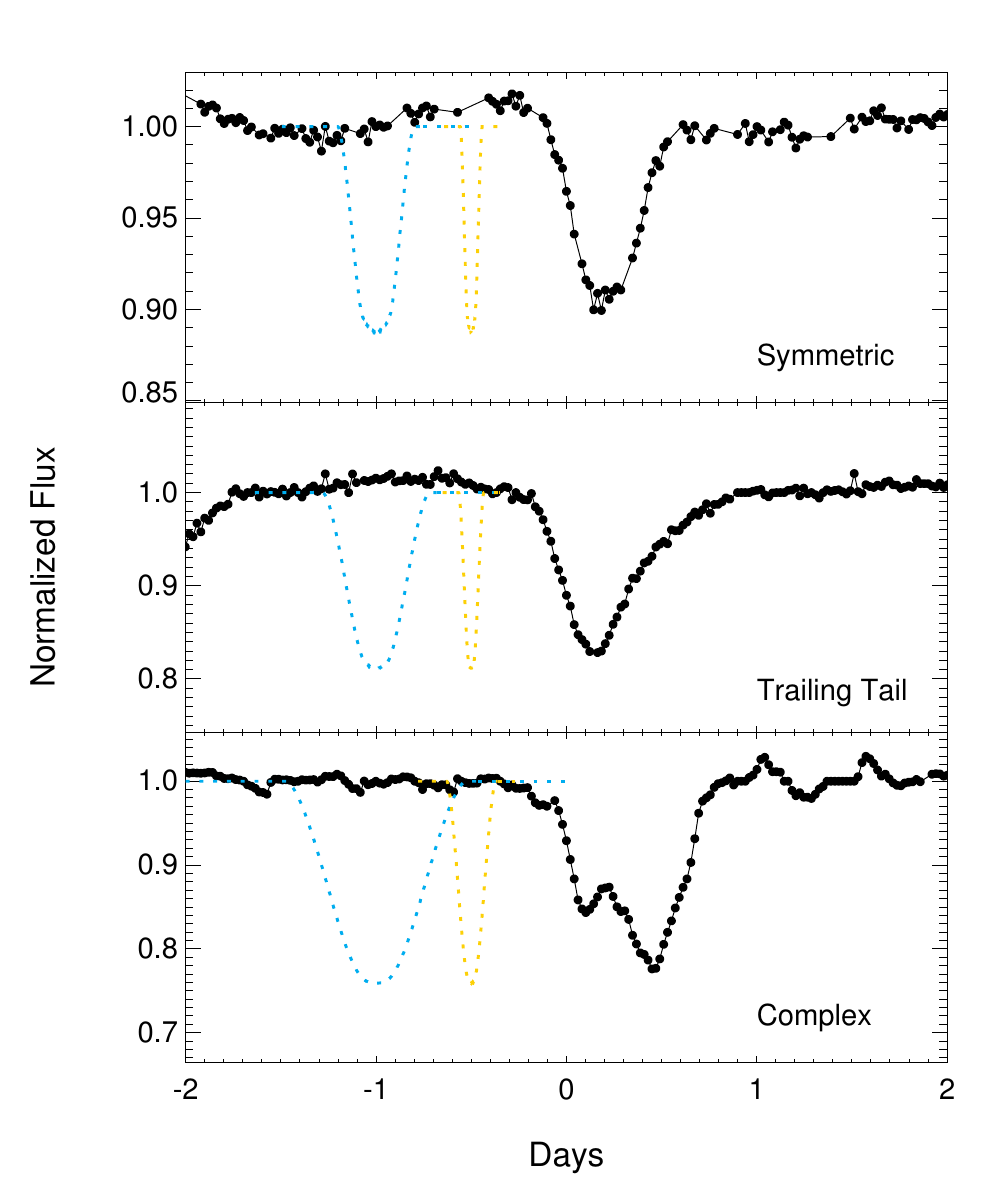}
\caption{\small Examples of the three main dip types seen in the K2/C2 light curves (black points). We also plot model planetary transits (offset for clarity) of the same depth to show their inconsistencies with the dips (see also Section 6). Yellow transit curves have durations based on circular orbits with periods equal to our $P_{\rm rot}$ value for the source, while blue transit curves have durations extended to match the FWHM of the dip. Model transits use the \cite{2004A&A...428.1001C} limb-darkening laws interpolated for the $T_{\rm eff}$ value of the star with an impact parameter of 0.5. Symmetric dips (top) resemble planet transits in shape, but require transit durations that are too long for the orbital period as well as transit depths that are too deep to be planetary. Asymmetric dips with lagging tails (middle) or complex structures (bottom) are clearly non-planetary.}
\label{fig-planets}
\end{centering}
\end{figure}
\capstartfalse

\subsection{Dipper Classification \label{sec-classification}}

The K2/C2 light curves of the dippers in our sample (Figure~\ref{fig-lcs}) show quasi-periodic or aperiodic dimming events that appear as discrete dips in flux with typical durations of $\sim$0.5--2.0 days and amplitudes of up to $\sim$40\%. Following \cite{2014AJ....147...82C} and \cite{2015A&A...577A..11M}, we define quasi-periodic dippers as those with dimming events that appear at periodic intervals but with varying shapes and depths, and aperiodic dippers as those with dimming events that appear stochastically and with varying shapes and depths. To classify the periodicity, we fit a hyperbolic secant function with time-varying dip depths to each light curve:
\begin{equation}
y = F_{o}- \sum_n 2D(t_{n}) \times \Big[  \exp\Bigl( \frac{t-t_{n}}{\tau} \Bigr) + \exp\Bigl(- \frac{t-t_{n}}{\tau} \Bigr)  \Big]^{-1},
\label{eqn-secant}
\end{equation}
\vspace{5mm}

where $t_{n}=t_{\rm o} + nP_{\rm rot}$. Here $F_{o}$ is the baseline flux level, $t_{\rm o}$ is the location of the first dip in the light curve, $n$ is the dip number, $P_{\rm rot}$ is the inferred rotation period, $\tau$ is the dip duration, and $D(t_{n})$ is the depth of dip $n$. We used the {\sc MPFIT} implementation of the Levenberg-Marquardt technique for $\chi^2$ minimization \citep{2009ASPC..411..251M} to find the best-fit parameters for each source while fixing $P_{\rm rot}$ to our derived value (Section~\ref{sec-identification}; Table~\ref{tab-properties}). 

As shown in Figure~\ref{fig-wrap}, this function can reasonably reproduce the dip occurrence, width, and depth for quasi-periodic but not aperiodic dippers. We found that some dippers (e.g., EPIC 203937317) have dips that appear episodic in their K2/C2 light curves, but then occur in phase when the light curve is phase folded to their $P_{\rm rot}$ value; we consider these dippers as quasi-periodic, as this implies that the dimming events occur at periodic intervals (though not at every interval). Table~\ref{tab-properties} gives the periodicity classification (quasi-periodic or aperiodic) for each dipper in our sample. 

Figure~\ref{fig-planets} shows examples of the three main dip types seen in the K2/C2 light curves, illustrating that the dips are clearly inconsistent with planetary transits.

 \subsection{Cluster Membership \label{sec-membership}}

The dippers in our sample are all members of either the Upper Sco or $\rho$ Oph star-forming regions (Table~\ref{tab-properties}), despite being selected only on their optical light curve properties. EPIC 203410665 (V* V896 Sco) and 203937317 (DoAr 24) are both well-known members of $\rho$ Oph that have been studied in the near-IR \cite[e.g.,][]{2009ApJ...703.1964F,2010ApJS..188...75M}. EPIC 205151387 is a well-studied member of Upper Sco whose disk has been detected with {\it Herschel} PACS \citep{2013A&A...558A..66M}, the Sub-millimeter Array \cite[SMA;][]{2008ApJ...686L.115C}, and the Atacama Large Millimeter Array \cite[ALMA;][]{2014ApJ...787...42C}. EPIC 203895983, 204630363, and 205519771 were spectroscopically confirmed as low-mass members of Upper Sco by \cite{2015MNRAS.448.2737R}. EPIC 204757338 and 204932990 were part of the \cite{2012ApJ...758...31L} sample of disk-bearing members in Upper Sco. EPIC 203343161 was classified as an Upper Sco member based on its photometry and proper motion \citep{2013MNRAS.431.3222L}. EPIC 204137184 was identified as a candidate low-mass member of Upper Sco, though based only on photometry \citep{2000AJ....120..479A}. 

\capstartfalse
\begin{figure*}
\begin{centering}
\includegraphics[width=17.5cm]{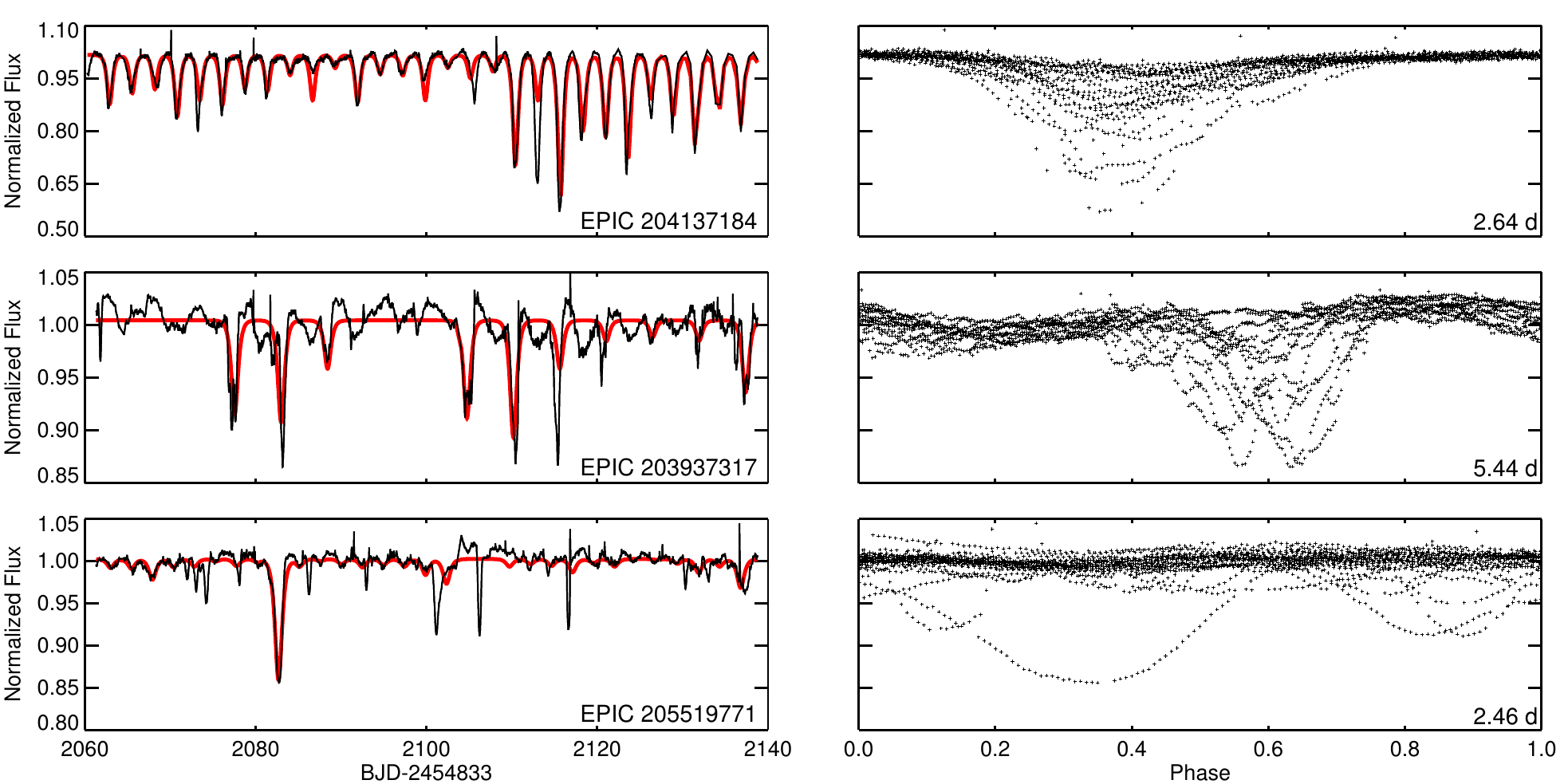}
\caption{\small Examples of quasi-periodic (top, middle) and aperiodic (bottom) dippers. Red lines show fits to the normalized light curves (left) using a hyperbolic secant function with time-varying dip depths (Equation~\ref{eqn-secant}; Section~\ref{sec-classification}). The function can reasonably reproduce dip occurrence, width, and depth for quasi-periodic but not aperiodic dippers. Phase-folded light curves (right) show dips occurring at similar phases for quasi-periodic but not aperiodic dippers. The $P_{\rm rot}$ values used for phase folding are given for reference (see also Table~\ref{tab-properties} \& Section~\ref{sec-identification}).}
\label{fig-wrap}
\end{centering}
\end{figure*}
\capstartfalse


\section{FOLLOW-UP OBSERVATIONS \label{sec-obs}}

We obtained multi-wavelength follow-up observations of our dippers, namely: optical and near-IR spectra (Section~\ref{sec-spectra}) to determine stellar properties and accretion states; sub-millimeter observations (Section~\ref{sec-sma}) to constrain disk dust and gas masses; and adaptive optics (AO) imaging (Section~\ref{sec-keck}) to search for close companions that could influence optical variations and/or disk evolution. We also utilized various literature and public data sources (Section~\ref{sec-aux}). These follow-up observations are described below and summarized in Table~\ref{tab-obs}.

\subsection{Spectroscopy\label{sec-spectra}}

We obtained optical spectra for our complete dipper sample using the Super-Nova Integral Field Spectrograph (SNIFS; \citealt{2002SPIE.4836...61A}; \citealt{2004SPIE.5249..146L}) at the University of Hawaii 2.2 m telescope atop Maunakea during 2015 May--July. These spectra have resolutions $R\approx900$ and signal-to-noise ratios ${\rm SNR}\gtrsim30$ per resolution element at $\gtrsim$5100 \AA{}. Details of our SNIFS data reduction can be found in \cite{2012ApJ...753...90M} and \cite{2013AJ....145..102L}. We used these spectra to estimate stellar properties (Section~\ref{sec-stellar}) and identify accretion states (Section~\ref{sec-accretion}). We performed spectral monitoring over several months (Table~\ref{tab-obs}) to study activity variability and account for changes in apparent effective temperature ($T_{\rm eff}$) due to variations in star spot coverage.

We used the Echellette Spectrograph and Imager (ESI) at the Keck II 10 m telescope atop Maunakea to obtain optical spectra of seven dippers in our sample on 2015 July 17. We used the instrument in cross-dispersed mode with a 0.5\as{} slit to achieve $R\approx8000$. Integration times of a few minutes yielded ${\rm SNR}\sim20$--90 in the Li {\sc I} 6708 \AA{} line (Table~\ref{tab-obs}). Spectra were reduced using the white dwarf standard EG 131 with the publicly available ESIRedux pipeline\footnote{\url{http://www2.keck.hawaii.edu/inst/esi/ESIRedux/index.html}} with dome flats for flat-fielding and arc lamps for wavelength calibration \citep{2003ApJS..147..227P,2009PASP..121.1409B}. Our ESI spectra were used to measure Li {\sc I} equivalent widths (Section~\ref{sec-stellar}) and search for spectroscopic binaries (Section~\ref{sec-binaries}).

We acquired near-IR spectra for eight dippers in our sample during 2015 May--July using the upgraded SpeX spectrograph \citep{2003PASP..115..362R} on the 3 m NASA Infrared Telescope Facility (IRTF) atop Maunakea. uSpeX spectra were taken in short cross-dispersed (SXD) mode using the 0.3\as{}$\times$15\as{} slit, covering 0.7 to 2.5 $\mu$m at $R\approx2000$. These spectra have ${\rm SNR}\gtrsim40$ in $K$ band (Table~\ref{tab-obs}). Basic reduction (bias, flat correction, extraction, etc.) was carried out with SpeXTool \citep{2004PASP..116..362C}. Flux calibration and telluric line removal were performed using A0V standards with Xtellcor \citep{2003PASP..115..389V}. Details on the observation, reduction, and extraction methods can be found in \cite{2015ApJ...804...64M}. We primarily used our uSpeX spectra to check for near-IR accretion signatures (Section~\ref{sec-accretion}).

\subsection{Sub-millimeter Observations\label{sec-sma}}

We obtained sub-mm observations with the SMA \citep{2004ApJ...616L...1H}, an interferometer on Maunakea consisting of eight 6 m antennas. Data were obtained during 2015 June--July for all dippers in our sample except EPIC 205151387, which was previously observed with the SMA by \cite{2008ApJ...686L.115C}. We used the SMA 1.3 mm band with the correlator configured to cover CO 2--1 line emission at 230.538 GHz. We used sub-compact or compact array configurations with baselines of $\sim$25 m or $\sim$50 m, respectively. The precipitable water vapor was typically $\sim$2.5 mm, corresponding to an atmospheric optical depth at zenith of $\sim$0.1 at 225 GHz. Each source was observed for 6 min in a cycle that was interleaved with 3 min integrations on quasars 1517-243 and 1625-254 for calibrating the variation of instrumental amplitude and phase with time. The tracks spanned hour angles of roughly $-2.5^{\rm h}$ to $+2.5^{\rm h}$ resulting in total integration times of $\sim$1 hr per source. The bandpasses were calibrated at the end of the tracks with 30 min integrations on 3C454.3 and the absolute flux scales were derived from observations of Titan. 

We used the MIR software package\footnote{\url{https://www.cfa.harvard.edu/~cqi/mircook.html}} to calibrate the data and the MIRIAD software package\footnote{\url{https://www.cfa.harvard.edu/sma/miriad}} for imaging. The close configuration of the array resulted in large beam sizes of roughly 8\as{}$\times$5\as{}. The achieved rms noise level for each source is given in Table~\ref{tab-obs}. We used our SMA observations to constrain disk dust and gas masses (Section~\ref{sec-mass}), which are important for classifying disk evolutionary states.

\subsection{Adaptive Optics Imaging\label{sec-keck}}

We obtained AO imagery for five dippers in our sample with the Near-infrared Imaging Camera (NIRC2) on the Keck II 10 m telescope atop Maunakea on 2015 June 22 using laser guide star AO \cite[LGS-AO;][]{2006PASP..118..297W,2006PASP..118..310V}. Imaging was done in $K^{\prime}$ band with the narrow camera, which has a pixel scale of $9.952\pm0.002$ mas pixel$^{-1}$ and an orientation angle of $0\fdg252\pm0\fdg009$ \citep{2010ApJ...725..331Y}. Weather conditions were photometric with seeing $<0.5^{\prime\prime}$. These NIRC2  images can reveal objects with $\Delta K^{\prime} \sim7$--8 mags at 1000 mas or $\Delta K^{\prime} \sim4$ mags at 100 mas. At the distance of Upper Sco and $\rho$ Oph, $\sim$100 mas corresponds to $\sim$15 AU, thus we used these data to search for close companions that could contribute to optical variations and/or influence disk evolution (Section~\ref{sec-binaries}).

\subsection{Archival Data\label{sec-aux}}

All dippers in our sample were observed by the Two Micron All-Sky Survey \citep[2MASS;][]{2006AJ....131.1163S}, supplying near-IR photometry in $JHK_{\rm S}$ bands. The dippers were also detected by the Wide-field Infrared Survey Explorer \cite[WISE;][]{2010AJ....140.1868W}, providing mid-IR photometry at 3.4, 4.6, 12, and 22 $\mu$m bands, denoted as W1 through W4. EPIC 203410665, 203937317, and 205151387 were observed with {\it Spitzer} IRS, providing mid-IR spectra covering the silicate emission feature at 10 $\mu$m. EPIC 205151387 was observed with {\it Herschel} PACS at 70, 100, and 160 $\mu$m bands \citep{2013A&A...558A..66M}. For EPIC 203937317, we were able to estimate its PACS 70 $\mu$m flux from a {\it Herschel} archive image. EPIC 205151387 was detected by ALMA at 880 $\mu$m \citep{2014ApJ...787...42C} and the SMA at 1.3 mm \citep{2008ApJ...686L.115C}. These archival data are summarized in Table~\ref{tab-phot}.


\section{ANALYSIS\label{sec-analysis}}

\subsection{Stellar Properties\label{sec-stellar}}

We estimated $T_{\rm eff}$ values for each dipper following the procedure in \cite{2013ApJ...779..188M} and \cite{2014MNRAS.443.2561G}. This method can reproduce interferometric $T_{\rm eff}$ measurements of nearby M dwarfs \citep{2012ApJ...757..112B} to $\simeq60$ K, but is untested on young ($\lesssim10$ Myr) stars. In short, we compared our SNIFS spectra to the CFIST grid of the BT-SETTL version of the PHOENIX stellar atmosphere models \citep{2013MSAIS..24..128A} with the relative abundances of solar elements as estimated by \cite{2011SoPh..268..255C}. We explored a grid of log {\it g} from 3.5 to 5.0 and $T_{\rm eff}$ from 2500 to 5000 K, but restricted metallicity and alpha abundance to solar values because large deviations from solar are unlikely for nearby, pre-main sequence stars. We fit and divided out a third-order polynomial in flux to capture the effects of reddening and imperfect flux calibration as well as a linear term in wavelength to correct for small errors in wavelength calibration and radial velocity offsets. The $T_{\rm eff}$ value for a given spectrum was taken from the best-fit model, with measurement errors typically $\lesssim$100 K. However, as a star rotates we observe different star spot distributions, which can change the derived $T_{\rm eff}$ value due to the cooler temperature of star spots relative to the surrounding photosphere. Thus for each dipper we obtained $T_{\rm eff}$ values for each SNIFS observation epoch (Table~\ref{tab-obs}), taking the weighted mean of these values as the final $T_{\rm eff}$ estimate reported in Table~\ref{tab-disk}. Errors were determined by adding in quadrature the standard error on the mean and the aforementioned typical measurement errors.  

Spectral subtypes were determined from spectral index measurements of molecular bands in our SNIFS spectra. Specifically, we applied correlations between spectral subtype and the depths of CaH and TiO molecular bands derived in \cite{2013AJ....145..102L}. Reddening and spectral subtype are highly degenerate when using only photometry, however the application of spectral index measurements is much more robust because changes in reddening cannot reproduce the depth of the molecular bands. Although spectral indices can be affected by the lower gravity of young objects, our estimated spectral subtypes reported in Table~\ref{tab-disk} are consistent with our derived $T_{\rm eff}$ values and accurate to $\pm$ one spectral subtype. 

We estimated reddening ($A_V$) by comparing our SNIFS spectra to a sequence of template M dwarf spectra from the nearby TW Hydrae Association (TWA) taken as part of a spectroscopic study of nearby young moving groups (Mann et al., in prep). The TWA members are within the local bubble where interstellar extinction is negligible \citep{2009MNRAS.397.1286A} and have ages similar to the dippers in our sample \cite[$\sim$10 Myr;][]{2013ApJ...762..118W,2014A&A...563A.121D}. Template spectra were taken with the same instrument and reduced with the same methods as the dipper spectra. We masked out regions with strong emission (e.g., \ha{} at 6563 \AA) or telluric absorption features (e.g., O$_2$ near 7600 \AA), then artificially reddened each template spectrum using an $R_V=3.1$ reddening law from \cite{1989ApJ...345..245C}. We marginalized over the template choice, $A_V$, and radial velocity. The reddened templates reasonably reproduced the dipper spectra with typical best-fit reduced $\chi^2$ of $\sim$1.6, however the strong degeneracy between template spectral subtype and $A_V$ resulted in larger errors. Figure~\ref{fig-degeneracy} illustrates the fitting procedure and final estimates on $A_V$ are given in Table~\ref{tab-disk}.

We measured equivalent widths of the Li {\sc I} line at 6708.0 \AA{} (EW$_{\rm Li}$) from our higher-resolution ESI spectra. The presence of Li {\sc I} absorption is an indicator of stellar youth ($<$50 Myr) in late-type stars because the element is rapidly convected into the stellar core where it is destroyed. We simultaneously fit two Gaussians, one centered on the Li {\sc I} line at 6708.0 \AA{} (our ESI resolution was too low to resolve the doublet) and the other centered on the nearby Fe {\sc I} line at 6707.5 \AA{}. We used the fitted Li {\sc I} line to estimate EW$_{\rm Li}$ values while accounting for any Fe {\sc I} contamination. Errors were determined using a Monte Carlo method: we used the standard deviation of the flux in two continuum regions flanking the lines to add Gaussian-distributed noise to the observed spectrum, then repeated the fitting procedure 100 times, taking the mean and standard deviation as our final EW$_{\rm Li}$ value and error, respectively. The EW$_{\rm Li}$ values reported in Table~\ref{tab-disk} are consistent with those found for M dwarfs in Upper Sco \cite[e.g., see Figure 5 in][]{2015MNRAS.448.2737R}.

\capstartfalse                                                           
\begin{deluxetable}{ccccccc}                                              
\tabletypesize{\footnotesize}                                           
\centering                                                               
\tablewidth{250pt}                                                         
\tablecaption{Follow-up Observations\label{tab-obs}}                  
\tablecolumns{6}                                                         
\tablehead{                              
   \colhead{EPIC ID}                                      
 & \colhead{SNIFS\textsuperscript{a}}                                                          
 & \colhead{ESI\textsuperscript{b}}                                                                                                      
 & \colhead{uSpeX\textsuperscript{c}}                                                          
 & \colhead{SMA\textsuperscript{d}}                                                         
 & \colhead{NIRC2\textsuperscript{e}}                                                                                                      
 }                                          
\startdata                                                               
203343161  &  4  &  25  &   40  &   1.9   &  N \\              
203410665  &  7  & ---   &  250 &   1.6   &  Y  \\  
203895983  &  7  & ---   &  185 &   1.5   &  Y  \\  
203937317  &  6  &  90  &   75  &   1.8   &  Y  \\  
204137184  &  5  &  30  &   ---   &   1.5   &  N \\  
204630363  &  5  & ---   &   ---   &   1.5   &  N \\  
204757338  &  5  &  20  &  110  &   1.4   &  N \\  
204932990  &  7  &  30  &  100  &   1.6   &  Y \\  
205151387  &  5  &  60  &  125  &   2.0    & N \\  
205519771  &  7  & 30  &  100  &   1.3   &  Y  
\enddata    
\tablenotetext{a}{Number of SNIFS observation epochs (Section~\ref{sec-spectra}).}
\tablenotetext{b}{SNR of ESI spectra in the Li {\sc I} 6708 \AA{} line (Section~\ref{sec-spectra})}
\tablenotetext{c}{SNR of uSpeX spectra in $K$ band (Section~\ref{sec-spectra}).}
\tablenotetext{d}{Flux rms noise (mJy) of SMA observations (Sections~\ref{sec-sma} \& \ref{sec-aux})}
\tablenotetext{e}{NIRC2 imagery (Section~\ref{sec-keck}): Y = observed, N = not observed.}
\end{deluxetable}                                                        
\capstartfalse

 \capstartfalse                                                           
\begin{deluxetable*}{clrrrrccrcc}                                              
\tabletypesize{\scriptsize}                                                                                                          
\tablewidth{500 pt}                                                         
\tablecaption{Stellar and Disk Properties \label{tab-disk}}                  
\tablecolumns{11}                                                         
\tablehead{                              
   \colhead{EPIC}                                      
 & \colhead{$T_{\rm eff}$\textsuperscript{a}}                                                   
 & \colhead{SpT\textsuperscript{b}}                                                   
 & \colhead{$A_{V}$\textsuperscript{c}}  
 & \colhead{EW$_{\rm H\alpha}$.\textsuperscript{d}}                                                          
 & \colhead{EW$_{\rm Li}$.\textsuperscript{e}}                                                          
 & \colhead{He {\sc I}\textsuperscript{f}}                                                          
 & \colhead{EW$_{\rm Pa\gamma}$\textsuperscript{g}}                                                          
 & \colhead{M$_{\rm dust}$\textsuperscript{h}}                                                         
 & \colhead{Acc.\textsuperscript{i}}                                                   
 & \colhead{Disk.\textsuperscript{j}}  \\
   \colhead{} 
 & \colhead{(K)}                                                                                
 & \colhead{}                                                                                
 & \colhead{(mag)}                                                                                
 & \colhead{(\AA)}                                                                                
 & \colhead{(\AA)}                                                                                
 & \colhead{}                                                                                
 & \colhead{(\AA)}                                                                                
 & \colhead{($M_{\oplus}$)}                                                                                
 & \colhead{}                                                                                
 & \colhead{}                                                                                
 }                                          
\startdata                                                               
203343161  &  3120$\pm$115  &  M5.5  & $0.09^{+0.13}_{-0.09}$   &  $23.4^{+10}_{-8.9}$   &  -0.46$\pm$0.03     &  C     &  0.6$\pm$0.3  &  $<3$  &  W/C   &  Ev \\              
203410665  &  4045$\pm$50    &  K7.0  &  $1.00^{+0.13}_{-0.35}$  &  $  1.1^{+1.9}_{-1.4}$  &  ---                           &  R     &  1.3$\pm$0.1  &  $<3$  &  W      &  F/(P)TD   \\  
203895983  &  3655$\pm$75    &  M2.5  &  $0.31^{+0.19}_{-0.06}$  &  $11.5^{+3.4}_{-6.1}$  &  ---                           &  R/B &  0.4$\pm$0.1  &  $<3$  &  W/C  &  F/(P)TD    \\  
203937317  &  4070$\pm$50  &  K7.5  &  $2.57^{+0.13}_{-0.16}$  &  $15.3^{+3.6}_{-3.2}$  &  $-0.39$$\pm$0.02 &  C     &  1.2$\pm$0.2  &   4       &  C       &  F   \\  
204137184  &  3210$\pm$115  &  M4.5  &  $0.41^{+0.09}_{-0.13}$  &  $  2.7^{+0.8}_{-1.0}$  &  $-0.60$$\pm$0.03 &  ---    &  ---                  &  $<3$  &  W      &  F    \\  
204630363  &  4070$\pm$50  &  K7.5  &  $0.81^{+0.13}_{-0.13}$  &  $24.2^{+11}_{-8.2}$   &  ---                           &  ---    &  ---                  &  15      &  C       &  F/(P)TD  \\  
204757338  &  3145$\pm$100  &  M4.5  &  $1.03^{+0.09}_{-0.06}$  &  $  4.3^{+1.3}_{-1.2}$ &  $-0.49$$\pm$0.03  &  R    &  0.0$\pm$0.1  &  7        &  W      &  F/(P)TD   \\  
204932990  &  3390$\pm$115  &  M3.5  &  $0.88^{+0.31}_{-0.09}$  &  $  6.9^{+8.6}_{-3.7}$ &  $-0.40$$\pm$0.04  &  R/B &  0.0$\pm$0.1  &  $<3$  &  W      &  F/(P)TD   \\  
205151387  &  3975$\pm$130  &  M1.0  &  $0.78^{+0.13}_{-0.39}$  &  $  6.9^{+2.4}_{-5.0}$ &  $-0.47$$\pm$0.02  &  B    &  1.0$\pm$0.1  &  7--9    &  W      &  F/(P)TD    \\  
205519771  &  3390$\pm$100  &  M3.5  &  $0.78^{+0.13}_{-0.34}$  &  $  2.5^{+1.1}_{-0.8}$ &  $-0.48$$\pm$0.04  &  R   &   0.1$\pm$0.1  &  $<3$  &  W      &  Ev     
\enddata    
\tablenotetext{a}{Effective temperature measured from our SNIFS spectra (Section~\ref{sec-stellar}).}
\tablenotetext{b}{Spectral subtypes estimated from our SNIFS spectra, accurate to $\pm$ one spectral subtype (Section~\ref{sec-stellar}).}
\tablenotetext{c}{Extinction estimated from our SNIFS spectra (Section~\ref{sec-stellar}; Figure~\ref{fig-degeneracy}).}
\tablenotetext{d}{Mean \ha{} equivalent widths with high/low ranges observed across our SNIFS spectral monitoring campaign (Section~\ref{sec-accretion}; Table~\ref{tab-obs})}
\tablenotetext{e}{Li {\sc I} equivalent widths measured from our ESI spectra (Section~\ref{sec-stellar}).}
\tablenotetext{f}{He {\sc I} accretion signatures seen in our uSpeX data: C = centered absorption, R = redshifted absorption, B = blueshifted absorption (Section~\ref{sec-accretion})}
\tablenotetext{g}{H {\sc I} Pa$\gamma$ equivalent widths measured from our uSpeX spectra (Section~\ref{sec-accretion})}
\tablenotetext{h}{Disk dust masses estimated from our SMA data; for EPIC 205151387, we give the range found in the literature (Section~\ref{sec-mass}).}
\tablenotetext{i}{Accretion state based on EW$_{\rm H\alpha}$ and spectral subtype \cite[criterion from][]{2003ApJ...582.1109W}: W = WTTS, C = CTTS. We use W/C to indicate sources that showed a large range of EW$_{\rm H\alpha}$ values during our spectral monitoring campaign and thus could be classified as WTTS or CTTS. }
\tablenotetext{j}{Disk type determined from IR color excesses (Section~\ref{sec-color}) and SEDs (Section~\ref{sec-sed}): F = Full, TD = Transition Disk, Ev = Evolved. We use F/(P)TD to indicate disks that were classified as ``full" by their IR colors but have SEDs indicative of (pre-) transition disks.}
\end{deluxetable*}                                                        
\capstartfalse

\subsection{Accretion Indicators\label{sec-accretion}}

Spectral accretion signatures are produced by shocked gas falling onto the star or encountering the stellar magnetosphere, making them useful indicators of inner gaseous disks and stellar evolutionary states. We first searched our SNIFS spectra for \ha{} emission from H {\small \sc I} at 6563 \AA{}, which is exhibited by all our dippers. However, certain levels of \ha{} emission can be emitted by young, late-type stars that are active but whose disks have dissipated and thus are no longer accreting. \cite{2003ApJ...582.1109W} found that for accreting CTTS, \ha{} equivalent widths (EW$_{\rm H\alpha}$) were $\gtrsim$10 \AA{} for K7--M2.5, $\gtrsim$20 \AA{} for M3--M5.5, and $\gtrsim$40 \AA{} for M6 and later spectral subtypes. We therefore obtained EW$_{\rm H\alpha}$ values from our SNIFS spectra following the procedure in \cite{2013AJ....145..102L}: we measured flux over a 14 \AA-wide region (6557.61--6571.61 \AA) relative to pseudo-continuum regions (6500--6550, 6575--6625 \AA) and calculated errors using the Monte Carlo method described in Section~\ref{sec-stellar}. Table~\ref{tab-disk} gives the average and  high/low ranges of the EW$_{\rm H \alpha}$ values for each dipper across all SNIFS observation epochs. These levels of \ha{} emission are more consistent with stellar youth rather than ongoing accretion: according to the criteria of \cite{2003ApJ...582.1109W}, only two dippers (EPIC 203937317, 204630363) consistently exhibited \ha{} emission at levels expected for accreting CTTS.

We searched our uSpeX spectra for near-IR accretion signatures, focusing on Pa$\gamma$ emission from H {\small \sc I} at 1.094 $\mu$m and He {\small \sc I} absorption at 1.083 $\mu$m. Pa$\gamma$ emission arises from magnetospheric accretion as with \ha. He {\small \sc I} absorption at 1.083 $\mu$m is thought to be particularly sensitive to inner disk flows due to the metastability of the transition's lower level: the line is known to exhibit redshifted absorption due to in-falling gas and/or blue-shifted absorption due to inner disk winds, both of which are believed to be powered by accretion \citep{2006ApJ...646..319E}. As summarized in Table~\ref{tab-disk}, four sources showed Pa$\gamma$ emission and six sources exhibited redshifted and/or blueshifted He {\small \sc I} absorption. Nevertheless, these lines are weak compared to those seen in accreting CTTS \cite[e.g.,][]{2006ApJ...646..319E}. This implies low accretion rates, although edge-on inclinations may also affect line profiles. Some dippers (EPIC 204757338, 204932990, 205519771) show redshifted and/or blueshifted He I absorption but lack Pa$\gamma$ emission and are classified as WTTS.

We checked for \oi{} forbidden emission at 5577 \AA{} and 6300 \AA{} in our SNIFS and ESI spectra. These lines are hallmarks of UV-photodissociated OH and H$_{2}$O gas in the surface layers of inner- to mid-disk regions around T Tauri-like stars \cite[e.g.,][]{2011ApJ...735...90G}. Only EPIC 204932990 showed weak emission in both lines with equivalent widths of $\sim$1--2 \AA{}. To check whether this emission could be due to airglow features from the Earth's atmosphere known to occur at the same wavelengths, we searched our collection of $\sim$2,000 SNIFS spectra of late-type stars from the \cite{2011AJ....142..138L} catalog. We found only 5 stars ($\sim$0.2\%) with equivalent widths $\gtrsim0.5$ \AA{}, making it very unlikely that the detected \oi{} emission from EPIC 204932990 was due to airglow. No dipper showed Ca II IRT emission in its optical spectrum.

\capstartfalse
\begin{figure}
\begin{centering}
\includegraphics[width=8.5cm]{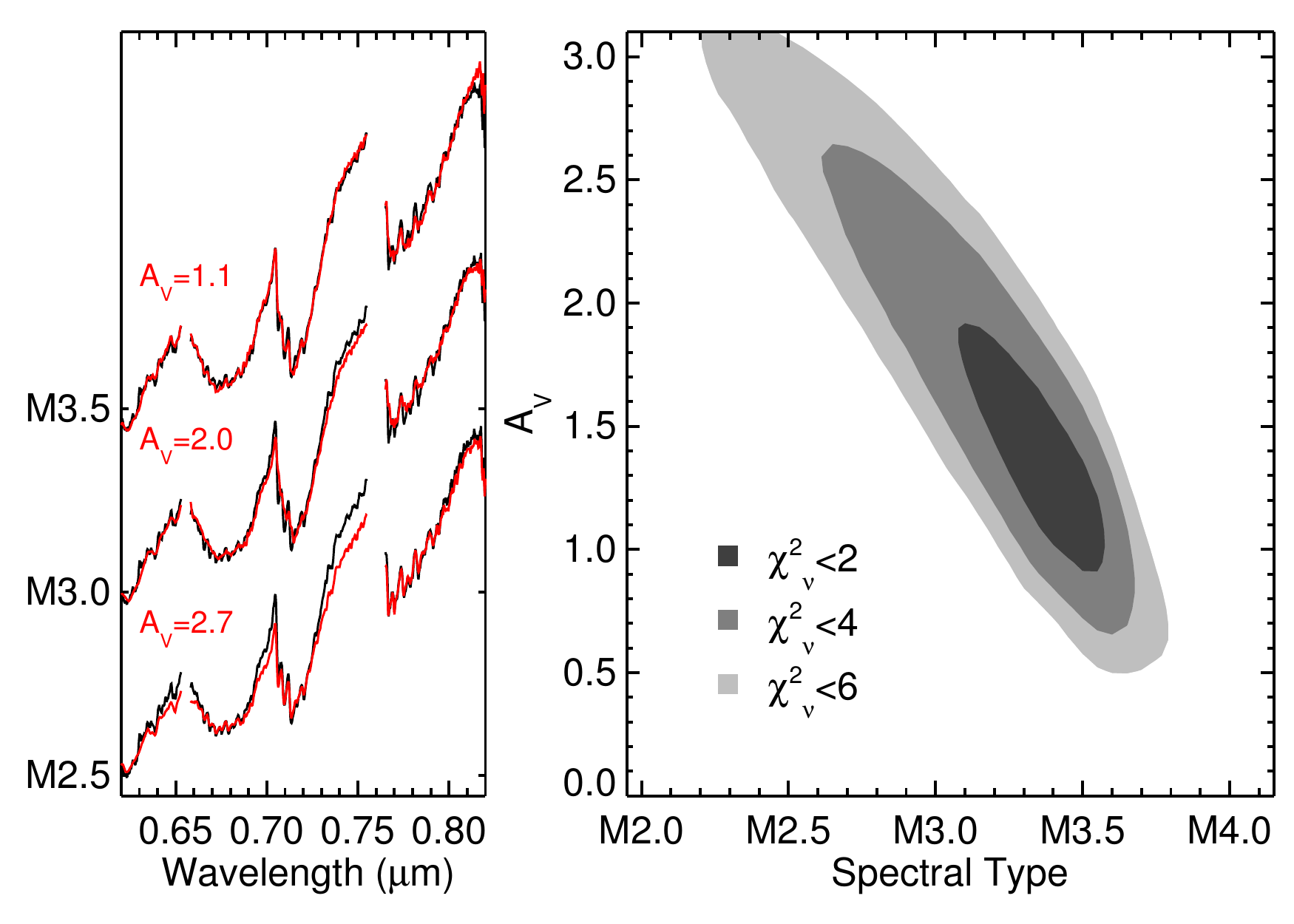}
\caption{\small Example of our extinction fitting technique. Left: SNIFS spectrum of EPIC 204932990 (black lines) compared to de-reddened template spectra from TWA (red lines); regions with strong emission or telluric absorption features are masked out. Right: The $\chi^2$ surface describing an expanded range of the fits shown on the left.}
\label{fig-degeneracy}
\end{centering}
\end{figure}
\capstartfalse

\subsection{Disk Dust \& Gas Masses \label{sec-mass}}

Sub-mm continuum emission can be used to estimate the total mass of dust in a disk. Because emission at sub-mm wavelengths from protoplanetary disks is typically optically thin, the continuum flux ($F_{\nu}$) is directly related to the total dust mass ($M_{\rm dust}$) as in \cite{1983QJRAS..24..267H}:  
\begin{equation}
M_{\rm dust}\approx\frac{F_{\nu}d^{2}}{\kappa_{\nu}B_{\nu}(T_{\rm dust})},
\label{eqn-mass1}
\end{equation}
where $B_{\nu}$ is the Planck function and we assume a characteristic dust temperature of $T_{\rm dust}=20$ K, which is the median found for Taurus disks \citep{2005ApJ...631.1134A}. The dust grain opacity, $\kappa_{\nu}$, is taken as 10 cm$^{2}$ g$^{-1}$ at 1000 GHz with an opacity power-law index $\beta=1$ \citep{1990AJ.....99..924B}. We assumed distances, $d$, of 140 pc for Upper Sco and 120 pc for $\rho$ Oph members. 

We detected only three dippers with the SMA at \textgreater 3$\sigma$ significance: EPIC 203937317 at $10.3\pm1.8$ mJy, EPIC 204630363 at $28.0\pm1.5$ mJy, and EPIC 204757338 at $12.5\pm1.4$ mJy. These sub-mm fluxes translate to dust masses of $\sim$4, 15, and 7 $M_{\oplus}$, respectively. EPIC 205151387 was previously observed with the SMA by \cite{2008ApJ...686L.115C} and ALMA by \cite{2014ApJ...787...42C}, who measured sub-mm fluxes corresponding to dust masses of 7--9 $M_{\oplus}$. The 3$\sigma$ upper limits for the undetected dippers translate to dust masses of $\lesssim$3 $M_{\oplus}$. 

Disk gas masses can be estimated from CO line emission. Converting integrated $^{12}$CO 2--1 line emission into total gas mass is complicated by the optical thickness of the line. Instead, our best constraints come from comparing CO isotopologue line emission to the grid of models in \cite{2014ApJ...788...59W}, who used these optically thin lines to predict total disk gas masses to within a factor of three. We did not detect $^{13}$CO or C$^{18}$O isotopologue line emission in our SMA data for any of the dippers at $\sim$40 mJy km s$^{-1}$ sensitivity. The $3\sigma$ upper limits only correspond to model disks from \cite{2014ApJ...788...59W} with gas masses $<1 M_{\rm Jup}$ (see their Figure 6), providing a rough upper limit to the dipper disk gas masses.

\subsection{Infrared Color Excess \& Variability \label{sec-color}}

For a typical young star hosting a disk, near-IR continuum emission arises from hot dust at $<$1 AU, while mid-IR emission probes warm dust in the inner tens of AU, and emission at far-IR and mm wavelengths traces cold dust in the outer disk regions. Thus IR excesses above the stellar photosphere can be used to infer radial disk structure and classify disk evolutionary states. ``Full" protoplanetary disks have not experienced significant clearing and thus exhibit strong excesses across the IR regime, indicating optically thick material throughout the disk. ``Transition" disks lack IR excesses at $\lesssim$10 $\mu$m but show strong excesses at longer wavelengths, reflecting inner cavities formed by disk clearing mechanisms (e.g., planet formation); a subset of ``pre-transition" disks exhibit near-IR excesses, indicating some optically thick material close to the star, but also mid-IR dips in their SEDs, indicating gaps in their disk structure \cite[see review in][]{2014prpl.conf..497E}. ``Evolved" disks have steadily decreasing IR excesses with increasing wavelength, reflecting their evolution toward optically thin disks. ``Debris" disks are composed of second-generation dust created from collisions among planetesimals; their dust masses are typically several orders of magnitude lower than those of full protoplanetary disks. Note that for cool M dwarf stars, near-IR emission from the disk is weak compared to that from the stellar photosphere, thus near-IR excess from the inner disk may not be readily apparent. 

 \cite{2012ApJ...758...31L} used IR color excesses relative to photospheric colors to classify disks in Upper Sco. They computed IR colors using {\it Spitzer} and WISE bands (which are dominated by dust emission) relative to $K$ band (which is dominated by stellar photosphere emission) then subtracted the IR color expected for a stellar photosphere of the same spectral subtype, as determined by fitting the sequence of diskless stars in Upper Sco (see their Figure 1). We calculated analogous IR color excesses using the W3 and W4 bands relative to the 2MASS $K_{\rm S}$ band: E($K_{\rm S}-{\rm W3}$) and E($K_{\rm S}-{\rm W4}$), respectively. Figure~\ref{fig-colors} plots these values against each other, indicating the different disk classifications for members of Upper Sco by color (Figure~\ref{fig-colors} is comparable to the bottom panel of Figure 2 in \citealt{2012ApJ...758...31L}). We over-plotted our dipper sample to assign initial disk types: the dippers appear to include full, evolved, and possibly transition disks. No dipper in our sample is classified as a debris or diskless system.

Some dippers in our sample exhibited mid-IR variability between their two W1/W2 observation epochs,  taken six months apart. Four dippers (EPIC 203410665, 203895983, 204137184, 204630363) exhibited $\sim$10--15\% increases/decreases in W1 and W2 flux over this time period, while the remaining dippers showed $\lesssim$5\% changes. (Pre-) transition disks are known to show so-called ``see-saw" mid-IR variability, where flux at shorter wavelengths increases/decreases while flux at longer wavelengths decreases/increases. This has been interpreted as changes in the height of the inner disk wall \cite[e.g.,][]{2011ApJ...728...49E,2012ApJ...748...71F} and could be related to the dipper phenomenon if the dimming events originate in the inner disk. Some young ($\sim$10--200 Myr) debris disks have also shown mid-IR variability on similar scales, interpreted as debris from planetesimal collisions \cite[e.g.,][]{2012Natur.487...74M}. For example, the young solar analog ID8 showed a $\sim$50\% brightening followed by a general decay over $\sim$1 year at both IRAC-1 and IRAC-2 bands \citep{2014Sci...345.1032M}. This could also be related to the dipper phenomenon, if the dimming events originate from clumps of debris in the disk.

\capstartfalse
\begin{figure}
\begin{centering}
\includegraphics[width=8.5cm]{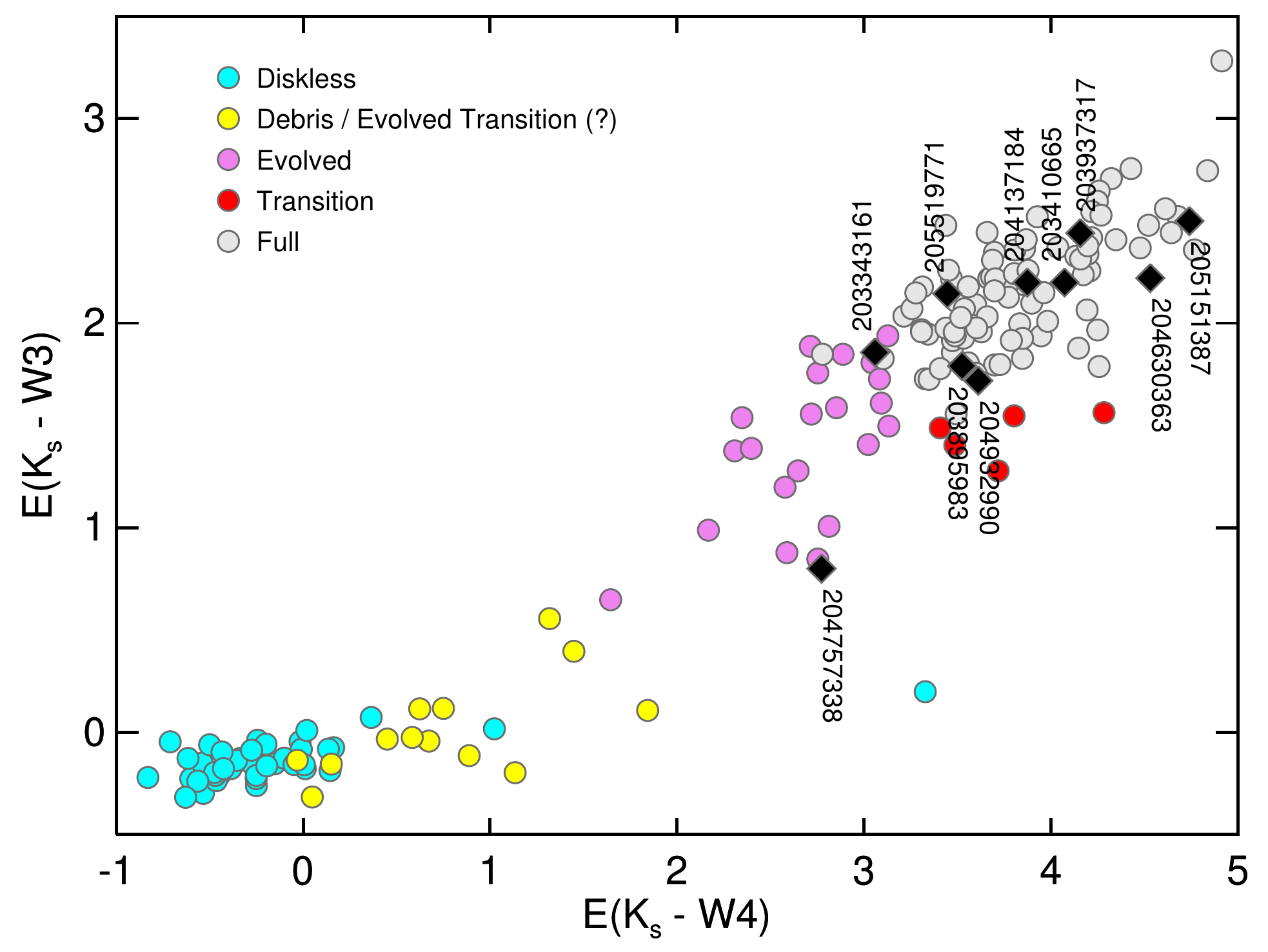}
\caption{\small IR color excesses used to classify disk types. The $x$ axis shows extinction-corrected $K_{\rm S}-{\rm W4}$ excess and the $y$ axis shows extinction-corrected $K_{\rm S}-{\rm W3}$ excess. Circles indicate late-type Upper Sco members, where colors specify their disk types assigned in \cite{2012ApJ...758...31L}. Black diamonds show the dippers in our sample.}
\label{fig-colors}
\end{centering}
\end{figure}
\capstartfalse

\subsection{Spectral Energy Distributions\label{sec-sed}}

SEDs for each dipper are shown in Figure~\ref{fig-sed} with photometry given in Table~\ref{tab-phot}. We derived the optical magnitudes from our SNIFS spectra by integrating over the revised filter profiles from \cite{2015PASP..127..102M}. We compared each SED to the expected stellar photosphere emission by plotting them against PHOENIX/CFIST model spectra with the same $T_{\rm eff}$ as the host star, corrected for extinction using our derived $A_{V}$ values (Table~\ref{tab-disk}) and the reddening law from \cite{1989ApJ...345..245C}. SEDs provide additional information that can be used to revise the initial disk classifications assigned by IR colors (Section~\ref{sec-color}). EPIC 203343161 and 205519771 have IR colors similar to evolved disks, and their homologously depleted SEDs support this classification. EPIC 203895983, 204630363, 204757338, 204932990, and 205151387 exist near the border between full and transition disks in Figure~\ref{fig-colors}, and their SEDs show little or no excesses at W1/W2 bands yet flat or rising excesses at W3/W4 bands, indicating they may be (pre-)transition disks. EPIC 203410665 and 204137184 are classified as full disks by both their IR colors and excesses at all WISE bands in their SEDs. EPIC 203937317 is classified as a full disk by its IR colors but shows no excess at W1/W2 bands, though this source is highly reddened. We summarize these disk types in Table~\ref{tab-disk}.

\capstartfalse
\begin{figure*}
\begin{centering}
\includegraphics[width=17.5cm]{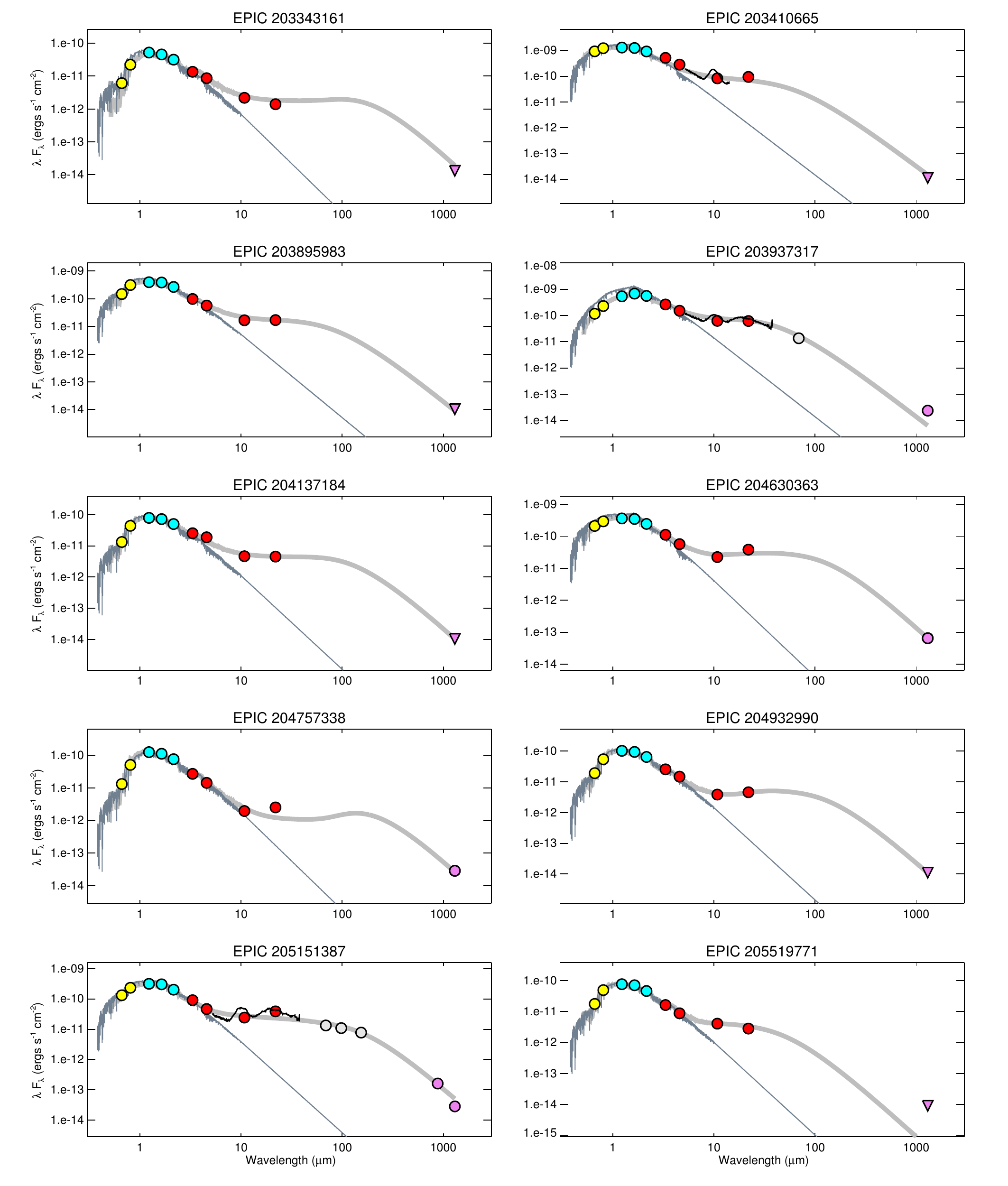}
\caption{\small SEDs of the ten dippers in our sample. Yellow points are $R_{\rm C}$ and $I_{\rm C}$ magnitudes derived from our SNIFS spectra, blue points are 2MASS $JHK_{\rm S}$ magnitudes, and red points are WISE 1--4 magnitudes. {\it Spitzer} IRS spectra (black lines) and {\it Herschel}/PACS magnitudes (gray points) are shown when available. Purple points are SMA 1.3 mm fluxes, where downward triangles indicate upper limits (Section~\ref{sec-sma} \& \ref{sec-aux}). EPIC 205151387 also has an ALMA observation at 880 $\mu$m from \cite{2014ApJ...787...42C} (Section~\ref{sec-aux}). Errors on the photometry points are all smaller than the symbols; the photometry is summarized in Table~\ref{tab-phot}. Dark gray curves show the expected stellar photosphere emission, derived from extinction-corrected PHOENIX/CFIST model spectra using the $T_{\rm eff}$ and $A_{V}$ values in Table~\ref{tab-properties} and the reddening law from \cite{1989ApJ...345..245C}. Thicker light gray curves show the SED models discussed in Section~\ref{sec-sed} with best-fit parameters given in Table~\ref{tab-sed}.}
\label{fig-sed}
\end{centering}
\end{figure*}
\capstartfalse

EPIC 203410665, 203937317, and 205151387 were observed by {\it Spitzer} IRS, providing mid-IR spectra that are over-plotted on their SEDs in Figure~\ref{fig-sed}. All three sources show the 10 $\mu$m silicate emission feature, thought to be produced by sub-micron crystalline grains at 300--500 K in the upper layers of the inner regions of flared disks \cite[e.g.,][]{1997ApJ...490..368C}. Following \cite{2007ApJ...659.1637S}, we measured the peak of the normalized flux at 10 $\mu$m (with respect to the continuum set at 1.0) finding values of 2.7, 1.8, and 2.1, respectively. These are higher than those of M dwarfs in the $\sim$4 Myr old cluster Tr 37, whose values are $\lesssim$1.5 \citep{2007ApJ...659.1637S}, though the silicate emission strength is a function of many factors such as stellar spectral type, disk viewing geometry, and grain properties.

We fit the dipper SEDs using models of emission from a star hosting a passive (i.e., non-accreting) circumstellar disk. For the SEDs, we used our SNIFS and uSpeX spectra plus the WISE and SMA photometry in Table~\ref{tab-phot}. The SNIFS and uSpeX spectra were stitched together using the overlap region at 0.85--0.95 $\mu$m, then placed on an absolute scale by constructing synthetic 2MASS $JHK_{\rm S}$ magnitudes and comparing them to observations. For modeling the host star, we interpolated on a grid of PHOENIX/CFIST model spectra with $T_{\rm eff}$ as the independent variable and assumed solar metallicity.\footnote{We re-fit $T_{\rm eff}$, rather than using the value derived from our optical spectra (Section~\ref{sec-stellar}), for self-consistency with the star-plus-disk model fit to the optical and IR data.} We assigned stellar masses and radii to each $T_{\rm eff}$ value using a 10 Myr isochrone from the Dartmouth Stellar Evolution Program \citep{2008ApJS..178...89D}, then computed $\log g$ from these values. A Gaussian weight for each model spectrum in the grid was computed; for fitting with an arbitrary value of $T_{\rm eff}$, a Gaussian-weighted sum of these spectra was used. Extinction of light from the star, $A_{\rm V}$, followed \cite{1989ApJ...345..245C} with $R_V = 4$. The model disks are composed of an optically thick but geometrically thin dust/planetesimal disk sandwiched between optically and geometrically thin upper layers of dusty gas \citep{1997ApJ...490..368C}. The former is illuminated at an oblique angle and thus relatively cool, while the dust in the latter is fully illuminated and thus hotter. The dust disk is heated both by the star and hotter upper layers. Emission from the inner layer was modeled as a perfect black body, while emission from the upper layers was modeled as a grey body. The disk is bounded by inner and outer radii ($a_{\rm in}$ and $a_{\rm out}$, respectively), and the inner radiating edge is limited also by the dust destruction temperature ($\sim$1700 K). The geometry of the finite angular extent of the star is taken into account.  The other parameters describing the disk are the inclination with respect to the LOS ($i_{\rm c}$, the co-inclination) and the upper layer emissivity ($\varepsilon$).  

The best-fit values of the two stellar and four disk parameters were found using the aforementioned {\sc MPFIT} routine and are reported in Table~\ref{tab-sed}.  A key issue with modeling the SEDs of cool, passive disks is that such disks emit little at $\lambda < 3 \mu$m and thus their only constraints are the WISE 1--4 magnitudes and an SMA detection or upper limit relative to the near-IR brightness. The number of free parameters (4) is thus comparable to the effective number of measurements (5), giving rise to degeneracy between the emissivity of the hot disk layer ($\varepsilon$) and disk co-inclination ($i_{\rm c}$). As a result of this degeneracy, there are acceptable fits over a range of inclinations and in some cases the best fits have high co-inclinations (i.e., face-on geometries).  Such geometries may be inconsistent with the interpretation that the dips are occultations from structures in the disk.  Possible reconciliatory explanations are that the disks are flared and/or warped. This would permit the inner regions of the disk to be seen edge-on while outer regions present more surface area to the observer.  This model also assumes a continuous disk with no gaps or other complex structures.  

 \capstartfalse                                                           
\begin{deluxetable}{lrrrrrr}                                              
\tabletypesize{\scriptsize}                                                                                                          
\tablewidth{240 pt}                                                         
\tablecaption{SED Modeling Results \label{tab-sed}}                  
\tablecolumns{7}                                                         
\tablehead{                              
   \colhead{EPIC}                                      
 & \colhead{$T_{\rm eff}$}                                                   
 & \colhead{$i_{\rm c}$}                                                   
 & \colhead{$A_{\rm V}$}  
 & \colhead{$a_{\rm in}$}                                                          
 & \colhead{$a_{\rm out}$}                                                           
 & \colhead{log($\varepsilon$)}  \\
    \colhead{}
 & \colhead{(K)}                                                   
 & \colhead{(deg)}                                                   
 & \colhead{(mag)}                                                   
 & \colhead{($R_{\ast}$)}                                                   
 & \colhead{($R_{\ast}$)}                                                   
 & \colhead{}                                                                                                         
 }                                          
\startdata                                                               
203343161  &  3050 &  1    &  1.3  &  9.6    &  1.5E+04   &   0.0 \\              
203410665  &  4660 &  15  &  1.0  &  1.1    &  7.4E+02   &  -1.3  \\  
203895983  &  3560 &  10  &  0.4  &  1.4    &  1.1E+03   &  -1.1  \\  
203937317  &  4200 &  4    &  3.0  &  1.0    &  7.2E+02   &  -0.6  \\  
204137184  &  3280 &  6    &  0.8  &  1.0    &  3.4E+03   &  -0.8  \\  
204630363  &  4160 &  17  &  1.1  &  24.4  &  3.3E+03   &  -1.2  \\  
204757338  &  2890 &  2    &  0.1  &  2.9    &  9.9E+03   &  -1.0  \\  
204932990  &  3280 &  55  &  0.7  &  38.8  &  1.2E+03   &  -1.6  \\  
205151387  &  3860 &  90  &  0.9  &  3.8    &  1.8E+03   &  -1.8  \\  
205519771  &  3300 &  52  &  0.5  &  8.1    &  1.7E+02   &  -1.6  
\enddata    
\end{deluxetable}                                                        
\capstartfalse

\subsection{Close Binaries\label{sec-binaries}}

Of the five dippers in our sample with NIRC2 AO images, three showed candidate close companions (Figure~\ref{fig-nirc}). EPIC 203410665 has a faint companion ($\Delta K_{\rm S}=1.8$ mag) with a separation of $\rho=0.896^{\prime\prime}\pm0.014^{\prime\prime}$ ($\sim$110 AU) and a position angle of ${\rm PA}=23\fdg07\pm0\fdg19$. The primary is a well-known member of $\rho$ Oph (V* V896 Sco), previously identified as a close binary in \cite{2015A&A...580A..88E}. EPIC 203895983 is a likely near-equal-mass binary ($\Delta K_{\rm S}=0.1$ mag) with $\rho=0.298^{\prime\prime}\pm0.001^{\prime\prime}$ ($\sim$40 AU) and ${\rm PA}=250\fdg41\pm0\fdg21$. EPIC 204932990 has a very faint candidate companion ($\Delta K_{\rm S}=5.6$ mag) with $\rho=1.080^{\prime\prime}\pm0.002^{\prime\prime}$ ($\sim$150 AU) and ${\rm PA}=232\fdg14\pm0\fdg07$. Errors were derived following \citealt{2015ApJS..216....7B}. Additional observations (e.g., multi-epoch astrometry) are needed to confirm that these candidate companions are not background objects, however the probability of unassociated objects within 1\as{} should be low as no other stars were seen in the 5\as{}$\times$5\as{} NIRC2 images. We also searched our ESI spectra for spectroscopic binaries; however, autocorrelations of the spectra after applying a high-pass filter showed no evidence for spectroscopic binaries for the seven dippers with ESI data

For the candidate close binaries shown in Figure~\ref{fig-nirc}, it is plausible that both components host disks given their similar ages. Thus the IR excess could be coming from one or both of the companions, as the WISE resolution is much larger than the binary separations. However, if the dipper phenomenon only occurs for disks seen nearly edge on, then it would be unlikely for both stars to contribute to the dipping events: \cite{1994AJ....107..306H} showed that coplanarity between the equatorial and orbital planes begins to diverge for binary systems with separations $>10$ AU (see their Figure 2). Moreover, our periodograms (Section~\ref{sec-identification}) only show one distinct period.

\capstartfalse
\begin{figure}
\begin{centering}
\includegraphics[width=7.5cm]{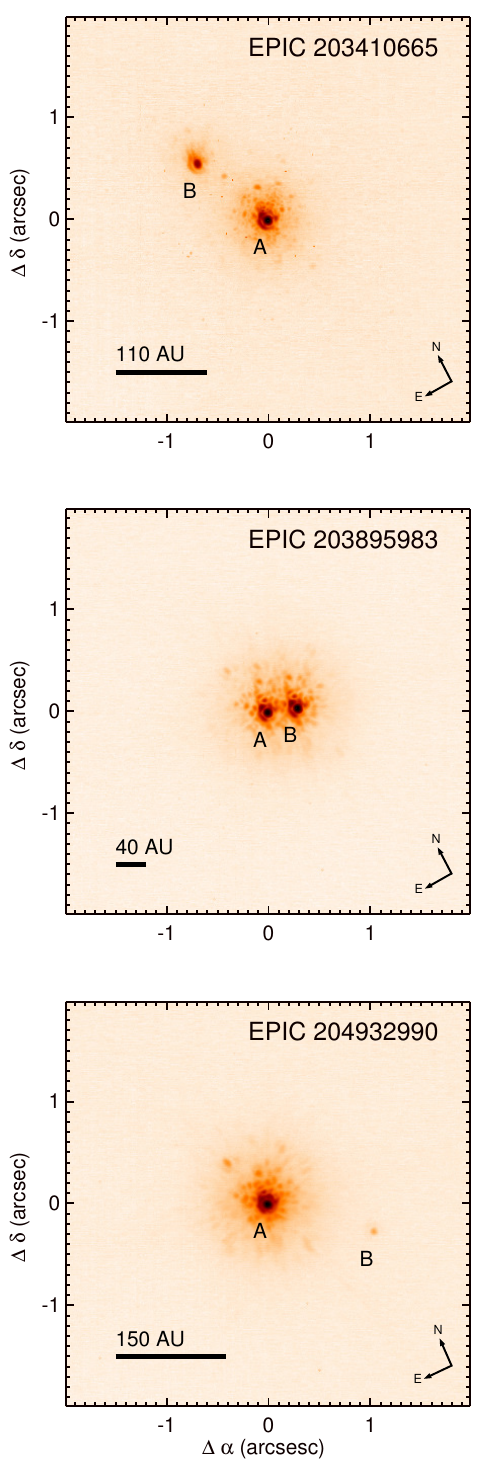}
\caption{\small NIRC2 AO images for the dippers in our sample with candidate companions. Primary (A) and secondary (B) components are labeled. Scalebars showing separations in AU are plotted for reference. North is up and East is left.}
\label{fig-nirc}
\end{centering}
\end{figure}
\capstartfalse


\section{Trends and Correlations\label{sec-trends}}

\subsection{Disk Structure \label{sec-structure}}

The dippers in our sample appear to host moderately evolved protoplanetary disks. For example, the dippers detected in the 1.3 mm continuum have dust masses of $\sim$4--15 $M_{\oplus}$ (Table~\ref{tab-disk}), which are an order of magnitude lower than typical protoplanetary disks at $\sim$10$^{2}$ $M_{\oplus}$, yet an order of magnitude higher than typical debris disks at $\sim$10$^{-1}$ $M_{\oplus}$ \citep[see Figure 3 in][]{2008ARA&A..46..339W}. Their upper limits on disk gas mass of $\sim$1 $M_{\rm Jup}$ also imply gas-to-dust ratios of \textless20--75, which are lower than the canonical ISM value of 100 \citep{1978ApJ...224..132B}; however, these gas-to-dust ratios are likely uncertain by at least a factor of three. Still, the weak accretion signatures exhibited by most of the dippers in our sample suggest accretion states between those of CTTS and WTTS (Section~\ref{sec-accretion}; Table~\ref{tab-disk}): \ha{} emission was typically consistent with stellar youth rather than ongoing accretion; only one dipper showed evidence of forbidden \oi{} emission; no dippers exhibited Ca II IRT emission; and the near-IR accretion signatures were weak compared to CTTS. 

Low disk masses are somewhat expected for the relatively old age of Upper Sco ($\sim$10 Myr). \cite{2013ApJ...771..129A} showed that disk masses in Upper Sco  are on average lower than those in the younger Taurus region ($\sim$2 Myr) by $\sim$2.5$\sigma$. Still, such low disk masses could be due in part to close companions. \cite{2012ApJ...751..115H} showed that mm luminosities of Taurus disks decrease by $\sim$5$\times$ for binary separations of $\sim$30--300 AU and by another $\sim$5$\times$ for separations of \textless30 AU. They interpreted this trend as being due to the tidal truncation of circumstellar disks in close stellar systems. The three dippers whose AO images revealed candidate companions in the $\sim$30--300 AU range have some of the lowest disk masses in our sample, yet the two other dippers with AO images have similarly low disk masses and appear to be single stars. Thus although close companions may have influenced the disks around some of our dippers, they cannot explain the low disk masses seen across our sample.

Moreover, half of our dippers show mid-IR dips in their SEDs, indicating they may host (pre-) transition disks that have developed gaps/holes in their disk structures (Table~\ref{tab-disk}). Our SED modeling (Section~\ref{sec-sed}) also suggests that EPIC 204932990 and 204630363 have inner holes, but that the remaining dippers have inner disks extending to within a few stellar radii. Note that our SED models do not account for pre-transition disks, which have gaps between inner/outer disk regions, as evident by the inability of the SED models to reproduce the mid-IR dips in some cases (e.g., 
EPIC 205151387; Figure~\ref{fig-sed}). 

\subsection{Correlations between disk and dip properties \label{sec-dipcolors}}

Figure~\ref{fig-dipcolors} shows E($K_{\rm S}-{\rm W2}$) vs. $D_{\rm dip}$ for all the late-type members of Upper Sco listed in \cite{2012ApJ...758...31L} that were also observed during K2/C2. $D_{\rm dip}$ is a measure of dip depth, while E($K_{\rm S}-{\rm W2}$) is an indicator of dusty material in the inner disk. $D_{\rm dip}$ was derived from the K2/C2 light curves as in Section~\ref{sec-identification}, and E($K_{\rm S}-{\rm W2}$) was calculated from the magnitudes listed in \cite{2012ApJ...758...31L} as in Section~\ref{sec-color}. The red line shows the $D_{\rm dip}$ cutoff that we used for identifying dippers (Section~\ref{sec-identification}); points above the cutoff include our dipper sample (black diamonds), additional dippers flagged by subsequent visual inspection of the K2/C2 light curves (orange points; see Table~\ref{tab-candidates}), and sources that did not meet our other dipper criteria and/or had corrupted light curves (gray points). 

Figure~\ref{fig-dipcolors} shows a positive correlation between $D_{\rm dip}$ and E($K_{\rm S}-{\rm W2}$) for the dippers (a Spearman rank test gives $\rho=0.51$ with a p-value of 0.01). The W2 band at 4.6 $\mu$m corresponds to a blackbody temperature of $\sim$600 K, which is roughly the temperature of dust grains orbiting at a few stellar radii around an early-M dwarf. Our SED modeling shows that most of our dippers likely have inner disks extending down to these small distances (Table~\ref{tab-sed}). We therefore interpret this correlation as evidence that the dips are related to inner disk material occulting the star. Indeed, this correlation with dip depth does {\it not} hold for W3 and W4 bands, which correspond to cooler material further out in the disk. Figure~\ref{fig-dipcolors} also shows that sources can have inner disks but not exhibit dips, which may be a geometric effect: if the dipper phenomenon requires disks with edge-on geometries, then stars with large W2 excess may not exhibit dips.

\capstartfalse
\begin{figure}
\begin{centering}
\includegraphics[width=8.5cm]{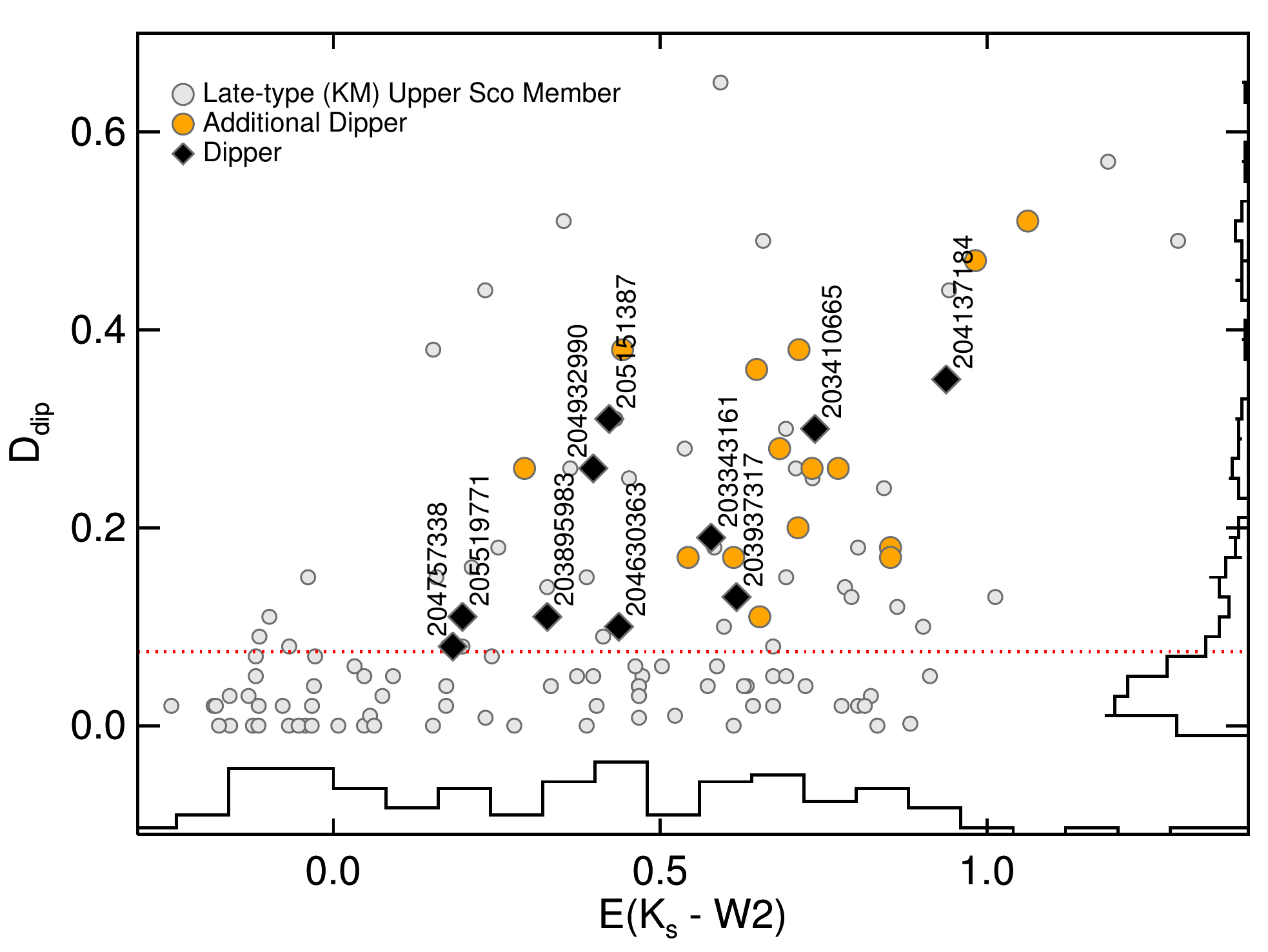}
\caption{\small W2 excess versus dip depth for late-type Upper Sco members in \cite{2012ApJ...758...31L} that were observed during K2/C2 (gray points). The red line shows our $D_{\rm dip}$ cutoff used for dipper identification (Section~\ref{sec-identification}). Our ten dippers (black diamonds; Table~\ref{tab-properties}) and the additional dippers (orange points; Table~\ref{tab-candidates}) are over-plotted. Histograms show the distributions of points in the $D_{\rm dip}-$ E($K_{\rm S}-{\rm W2}$) plane projected onto the $D_{\rm dip}$ and E($K_{\rm S}-{\rm W2}$) axes, respectively.} 
\label{fig-dipcolors}
\end{centering}
\end{figure}
\capstartfalse

 \capstartfalse                                                           
\begin{deluxetable}{ll}                                              
\tabletypesize{\footnotesize}                                           
\centering                                                               
\tablewidth{150 pt}                                                         
\tablecaption{Additional Dippers$^{\dagger}$ \label{tab-candidates}}                  
\tablecolumns{2}                                                         
\tablehead{                              
   \colhead{EPIC}                                      
 & \colhead{2MASS}                                                          
 }                                          
\startdata                                                               
203429083 & 15570350-2610081  \\
203824153 & 16285407-2447442  \\
203843911 & 16262367-2443138  \\
203850058 & 16270659-2441488  \\
203862309 & 16274270-2438506  \\
203969672 & 16270907-2412007  \\
203995761 & 16281673-2405142  \\
204107757 & 15560104-2338081  \\
204211116 & 16214199-2313432  \\
204329690 & 16220194-2245410  \\
204449274 & 16222160-2217307  \\
204489514 & 16030161-2207523  \\
204530046 & 16105011-2157481  \\
204864076 & 16035767-2031055  \\
205068630 & 16111095-1933320  
\enddata    
\tablenotetext{$\dagger$}{These targets were found by subsequent visual inspection of the K2/C2 light curves and will be the subject of future follow-up observations.}
\end{deluxetable}                                                        
 \capstartfalse

\subsection{Correlations with Stellar Properties\label{sec-latetype}}

The ten dippers in our sample are all late-K or M dwarf stars (Table~\ref{tab-disk}). This is unlikely to be the result of the community-based selection of the K2/C2 sample: of the $\sim$13,000 long-cadence targets, $\sim$55\% have $K_{\rm p}-J>1.8$, which is approximately the late-K and M dwarf regime. Thus the probability of picking ten random stars and having them all be late-K or M dwarfs is only $\sim$0.1\%. Moreover, $\sim$80\% of the early-type and $\sim$95\% of the late-type members of Upper Sco (based on the samples from \citealt{2012ApJ...758...31L} and \citealt{2015MNRAS.448.2737R}) were observed in K2/C2, making it unlikely that all the dippers would be cool dwarfs by chance.  

Rather, this bias toward late-type stars could result from the dipper phenomenon being related to circumstellar material, as disks persist significantly longer around low-mass stars compared to high-mass stars \cite[e.g.][]{2015A&A...576A..52R}. At the $\sim$10 Myr age of Upper Sco, the disks around early-type stars have experienced significant clearing: for example, \cite{2012ApJ...758...31L} found that the fraction of Upper Sco members with inner primordial disks is $\lesssim$10\% for early-type (BAFG) stars, but increases to $\sim$25\% for late-type (M5--L0) stars.

Furthermore, the bias toward late-type stars could imply that the dips are related to circumstellar material located at a specific distance from the star where a given temperature is reached. The lower stellar luminosities of M dwarfs mean that such temperature-dependent disk components exist closer to the star, where transiting orbits are more likely. Indeed, we find a positive correlation between $P_{\rm rot}$ and $T_{\rm eff}$ in our dipper sample: the cooler systems have $P_{\rm rot}\sim2$--3 days, while the hottest systems have $P_{\rm rot}\sim4$--10 days (a Spearman rank test gives $\rho=0.77$ with a p-value of 0.01). These inferred rotation periods correspond to dust grains at $\sim$400--600 K, which is where the W2 band peaks and we observe a correlation with dip depth (Figure~\ref{fig-dipcolors}), providing further support that the dipper phenomenon arises from the inner disk.


\section{Proposed Mechanisms \label{sec-mechanisms}}

Figure~\ref{fig-planets} shows representative examples of the three main dip types seen in the K2/C2 light curves (symmetric, trailing tail, complex). We compare these dips to model planetary transits in order to highlight their differences. Although the symmetric dips have similar shapes to transits of giant planets with short orbital periods around small stars, their dip durations and depths are both too large to be planetary. Similarly, although the dips with trailing tails resemble occultations by exo-comets \cite[e.g.,][]{1999A&A...343..916L}, their depths are several orders of magnitude deeper than expected. Below we discuss three possible mechanisms to explain the dipper phenomenon, all of which are related to dusty circumstellar material passing through our LOS to the star. No single mechanism can explain our entire dipper sample alone, yet collectively these mechanisms can account for the range of observed dipper properties; given the diversity of the dipper light curves, it may be that the dippers are produced by different mechanisms that correspond to different stages of disk evolution.

\subsection{Occulting Inner Disk Warps \label{sec-warps}}

Occultations of a star by non-axisymmetric structures in the inner disk could explain the dipper phenomenon, if the disk is seen nearly edge on. Given the amplitudes of the observed dips, the occulting material must have a vertical scale comparable to the size of the star.  Assuming the material is orbiting with a Keplerian period equal to the inferred stellar rotation period, $P_{\rm rot}$, (i.e., near the co-rotation radius) the co-inclination must be $\lesssim$$27 R_{\ast} P_{\rm rot}^{-2/3}M_{\ast}^{-1/3}$ degrees for an occultation to occur, where $R_{\ast}$ and $M_{\ast}$ are in solar units and $P_{\rm rot}$ is in days. This equates to a maximum co-inclination of $\sim$9 degrees for an M dwarf of $R_{\ast}=0.5~R_{\odot}$ and $M_{\ast}=0.5~M_{\odot}$ with $P_{\rm rot} = 2.5$ days. This means that a favorable geometry for occultation will occur $\sim$16\% of the time. However, of the $\sim$135 Upper Sco members with ``full" disks that were observed during K2/C2 (based on samples from \citealt{2012ApJ...758...31L} and \citealt{2015MNRAS.448.2737R}), we found only $\sim$10 ($\sim$7\%) to be dippers. 

One previously proposed explanation for the quasi-periodic and aperiodic dippers seen in the CTTS populations of young star-forming regions is occultation by an inner disk warp near the co-rotation radius, where the warp in the accretion disk arises from dynamical interactions with an inclined magnetosphere. In particular, \cite{2015A&A...577A..11M} explained the quasi-periodic and aperiodic dippers in the young ($\sim$3 Myr) NGC 2264 star-forming region as resulting from stable and unstable accretion regimes, respectively. In stable accretion regimes, the disk warp occults the star with each rotation, causing regular dips in the light curve \citep{1999A&A...349..619B}. In unstable accretion regimes \citep{2013MNRAS.431.2673K}, stochastic occultations of the star occur when dust is lifted above the mid-plane near the base of an accretion column that passes along our LOS. \cite{2015A&A...577A..11M} found that some dippers in NGC 2264 switched between stable and unstable accretion regimes, exhibiting both quasi-periodic and aperiodic dimming over timescales of a few years.

This occulting inner disk warp scenario can explain the quasi-periodic (e.g., EPIC 203937317) and aperiodic (e.g., EPIC 203410665) dippers in our sample with clear signs of accretion. To check the plausibility of this scenario for these dippers, we estimated the lower limit on the semi-major axis of a disk warp, $a_{\rm warp}$, by setting the Keplerian orbital period to the dip duration, $\tau$, as shorter rotation periods would imply that the warp extends further than the distance around its orbit. For a star with $M_{\ast} = 0.5~M_\odot$ and $R_{\ast}=0.5~R_\odot$, this yields $a_{\rm warp}>4$, 7, and 11~$R_{\ast}$ for $\tau=0.5,$ 1.0, and 2.0~days. These values are comparable to the co-rotation radii derived for M dwarf dippers in NGC 2264 by \cite{2015A&A...577A..11M}. This implies that disk warps extending around most of their orbit, and producing dips lasting up to a couple of days, will exist near the co-rotation radius, as expected for the occulting disk warp scenario. Nevertheless, this scenario is difficult to apply to the majority of dippers in our sample, which have weak accretion signatures and low disk masses.

\subsection{Vortices Produced by the Rossby Wave Instability\label{sec-rwi}}

Non-axisymmetric structures in the inner disk can also arise from the Rossby Wave Instability (RWI). Rossby waves grow around extrema in the inverse potential vorticity of disks \citep{2014FlDyR..46d1401L} until they evolve into vortices \citep{2010A&A...516A..31M} that are stabilized by local pressure maxima \citep{2010ApJ...725..146P}. These extrema can occur at the boundaries between turbulent, magnetized, and accreting regions in the disk and ``dead zones'' where flows are laminar and viscous transport is low \citep{2006A&A...446L..13V}. \cite{2015AJ....149..130S} proposed such vortices as a possible explanation for a subset of shallow, short-duration, periodic dippers in NGC 2264 (their ``Model 3'') but concluded that the degree of obscuration was insufficient to explain the dip amplitudes. We revisit this problem below, arguing that RWI-driven vortices can explain the quasi-periodic dippers in our sample with weak accretion signatures (e.g, EPIC 204757338), which are not readily explained by the disk warp scenario (Section~\ref{sec-warps}). 

The 3D gas dynamical simulations by \cite{2013A&A...559A..30R} numerically confirm the analytic predictions of \cite{2009A&A...498....1L} and \cite{2013ApJ...765...84L} that instabilities initially proceed rapidly at large azimuthal wave numbers but eventually the vortices merge. Only vortices with large (\textgreater6) aspect ratios in the radial-azimuthal plane survive breakdown by the elliptic streamline instability, and these vortices will typically span 1 radian of the orbit \cite[e.g., Figure 3 in][]{2013A&A...559A..30R}. These characteristics of RWI-generated vortices explain the single mode of the quasi-periodic dippers as well as their dip depths and durations, which require that the occulting structures cover a significant fraction of their orbit. Indeed, the quasi-periodic dips seen in our sample appear to span roughly 1 radian (Figure~\ref{fig-wrap}). Moreover, RWI-generated vortices extend vertically across the scale height $H$ of the disk, so if the vortices are optically thick then they will obscure a fraction $H/R_{\ast}$ of the star for nearly edge-on inclinations. For molecular hydrogen at $\sim$600 K orbiting at 2.5 days around a 0.4 $R_{\odot}$ star, this equates to a dip depth of 15\%, which is characteristic of our dippers. 

Such vortices can be efficient traps of dust, enhancing local surface densities by 1--2 orders of magnitude, although high dust concentrations may ultimately cause instabilities \citep{2014ApJ...795L..39F,2015MNRAS.450.4285C}. The absolute dust mass required to produce a dip is modest: the minimum mass of dust with diameter $d$ required to produce a dip of depth $f$ is $2\pi fd R_*^2 \rho/3$. For a 10\% dip, this equates to $\sim$$10^{17}$~kg (the mass of a small asteroid) for mm-sized grains with bulk densities of 3 g cm$^{-3}$.  Assuming the vortex spans 1 radian of the orbit and has an aspect ratio of 6, this corresponds to a surface density of $10^{-2}$ g cm$^{-2}$.  For comparison, a few $M_{\oplus}$ of dust (based on our sub-mm observations) spread uniformly over a disk of radius $\sim$5 AU (based on our SED fitting) has a surface density of order 1 g cm$^{-2}$.

However, one issue with the vortex model is explaining how the amplitude and shape of the dips change between one orbit and the next, as simulations generally show that the vortices persist for 100s of orbits. One possibility is that, because of disk flaring, we only observe dips produced by vortices extending well above the disk mid-plane, as the LOS to the disk mid-plane would be obscured by the disk itself \citep{2013A&A...559A..30R}. Another issue is that the locations of these vortices would correspond to disk temperatures of $\sim$400--700 K, according to the inner disk edges derived from our SED modeling (Section~\ref{sec-sed}). These temperatures are well below the vaporization temperature of Na and K ($\sim$1000 K) where the disk gas is expected to become electrically conducting and therefore susceptible to MRI-driven turbulence.  Because of this transition, the distance in the disk where temperatures reach $\sim$1000 K is a candidate dead zone edge where vortices are likely to emerge. Still, mid-plane temperatures may be higher if there is a heat source internal to the disk, such as the viscous dissipation of the gravitational energy of accretion.

\subsection{Transiting Circumstellar Clumps \label{sec-clumps}}

The disk warp (Section~\ref{sec-warps}) and RWI-generated vortex (Section~\ref{sec-rwi}) scenarios are difficult to reconcile with the aperiodic dippers in our sample with weak accretion signatures and low disk masses (e.g., EPIC 205519771). This is because disk warps/vortices near the co-rotation radius should produce regular dimming events every few days. Unstable accretion (related to disk warps; Section~\ref{sec-warps}) is also unlikely for these sources and cannot explain the widely separated dimming events. Thus we explore an alternative scenario where aperiodic dimming events result from single transits of dusty clumps of circumstellar material embedded at a few AU in the disk.

Protoplanetary disks are expected to be intrinsically clumpy because the gas distribution can exhibit low-level inhomogeneities that are amplified in the dust distribution \cite[e.g.,][]{2013A&A...550L...8B}. Concentrations of particles at specific locations can lead to streaming instabilities, which further enhance over-densities \citep[e.g.,][]{2007Natur.448.1022J}. Gravitational instabilities may then form $\sim$1000 km-diameter planetesimals \citep{2011A&A...529A..62J}. These processes have received much attention in numerical studies, as they can explain how dust grows into planetesimals, overcoming barriers of radial drift and bouncing \cite[e.g.,][]{2012A&A...540A..73W}. Here we consider whether these physical processes can result in disks that are intrinsically clumpy at the levels required to explain the aperiodic dippers.

If we assume an equatorial transit and a circular orbit with speed $v_{\rm orb}$, we can write the clump transit time across the host star (i.e., the dip duration) as $\tau = 2(R_c+R_*)/v_{\rm orb}$ days, where $R_{c}$ is the radius of the clump in solar radii and $R_{\ast}$ is the radius of the host star in solar radii. From this, and the assumption that the clump masses are much smaller than the stellar mass, we can derive an expression for the clump size:
\begin{equation}
R_{c} \approx 1.85 \, \tau \, \Big(\frac{M_{\ast}}{a}\Big)^{1/2} - R_{\ast},
\label{eqn-size}
\end{equation}
where $a$ is the semi-major axis of the clump orbit in AU and $M_{\ast}$ is the mass of the host star in solar masses. The correlation between clump size and semi-major axis from Equation~\ref{eqn-size} is illustrated in Figure~\ref{fig-clumps} for a star of $M_{\ast} = 0.5~M_\odot$ and $R_{\ast}=0.5~R_\odot$ with dip durations of $\tau=0.5$, 1.0, and 2.0 days.

We can place lower limits on $a$ and $R_{c}$ based on the nature of the aperiodic dimming events, which imply that the clumps have orbital periods longer than the 80-day K2/C2 observing campaign. Combining the requirement of $P>80$ days with Kepler's third law implies a lower limit to the semi-major axis of the orbiting clumps:
\begin{equation}
a > \left(GM_*\right)^{1/3} \left(\frac{P=80~{\rm days}}{2 \pi}\right)^{2/3} \approx 0.36 \, M_{\ast}^{1/3} ~{\rm AU}.
\label{eqn-axis}
\end{equation}
A lower limit to $R_{c}$ can be inferred from the minimum dip depth; using $D_{\rm dip}$ (Table~\ref{tab-properties}) as a measure of minimum dip depth, the lower limit to the clump size is: 
\begin{equation}
R_{c} > R_{\ast} \, D_{\rm dip}^{1/2}.
\label{eqn-sizelimit}
\end{equation}
We show these observational restrictions on $a$ and $R_{c}$ as gray regions in Figure~\ref{fig-clumps}. Note that the limit on $R_{c}$ corresponds to a dip depth of $D_{\rm dip}\sim0.1$ and that these limits only apply to the aperiodic dippers in our sample.

\capstartfalse
\begin{figure}
\begin{centering}
\includegraphics[width=8.5cm]{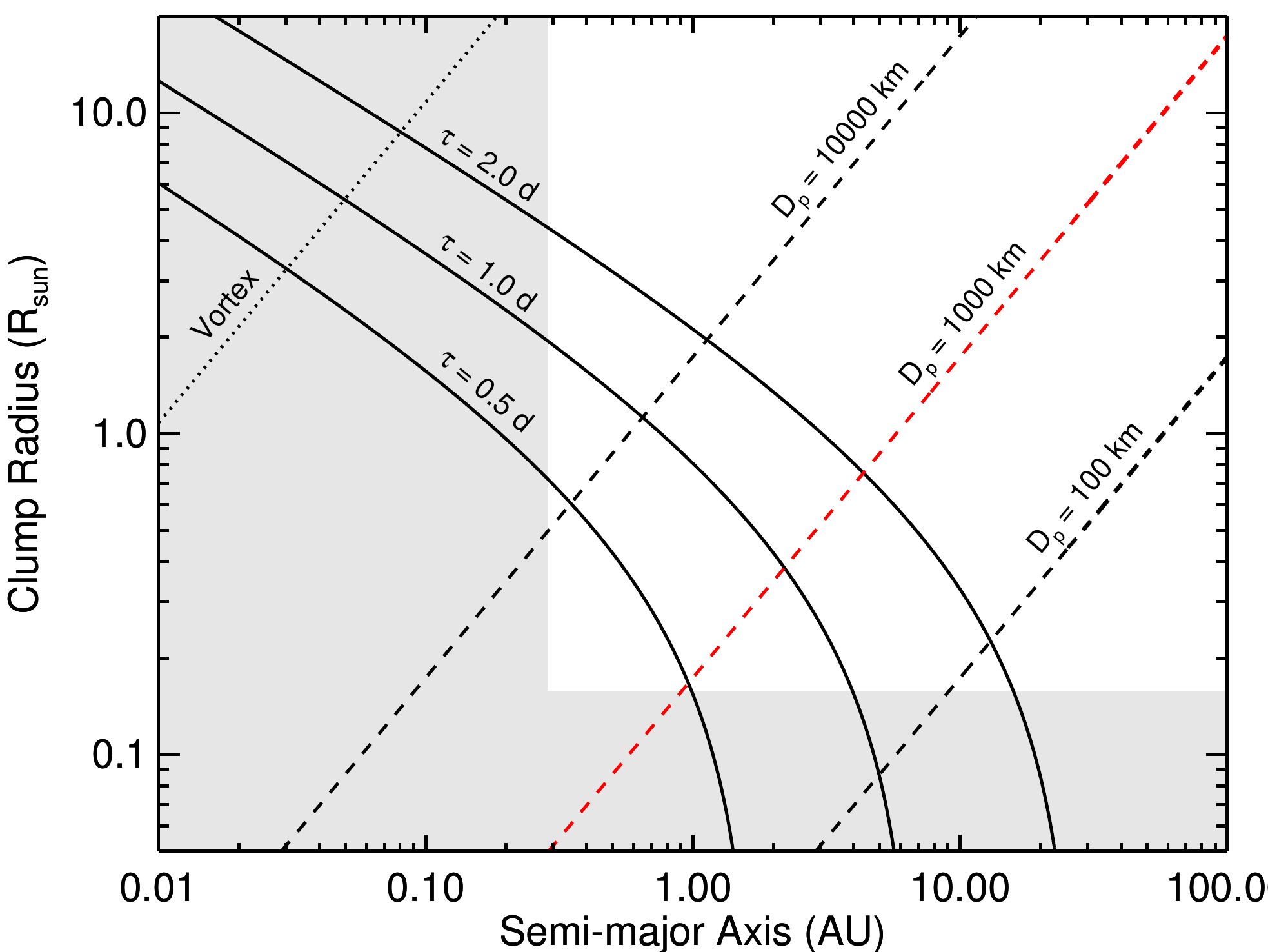}
\caption{\small Relations between clump size ($R_{c}$) and semi-major axis ($a$). Dashed lines show the Hill spheres of planetesimals with different diameters, while solid lines show contours of constant dip durations; both assume an M dwarf host star with $M_{\ast} = 0.5M_\odot$ and $R_{\ast}=0.5R_\odot$ (Equation~\ref{eqn-size}). Gray regions show forbidden values of $a$ and $R_{c}$ for the transiting clump scenario from Equations~(\ref{eqn-axis}) and ~(\ref{eqn-sizelimit}), respectively. The dotted line shows an example of an RWI-generated vortex spanning 1 radian around its orbit for comparison.}
\label{fig-clumps}
\end{centering}
\end{figure}
\capstartfalse

Could the transiting clumps be related to the $\sim$1000 km planetesimals formed via streaming instabilities? Although the planetesimals are too small to cause the observed dips, their Hill spheres may provide suitably sized obscuring clumps. Figure~\ref{fig-clumps} shows the Hill spheres of planetesimals with diameters of $D_{\rm p}=100$, 1000, and 10,000 km assuming bulk densities of 3 g cm$^{-3}$. The Hill spheres of $\sim$1000 km planetesimals orbiting at a few AU can produce the observed dip durations within our constraints on $R_c$ and $a$.

One concern with the transiting clump scenario is the required number of $\sim$1000 km planetesimals, if the dips are indeed independent from each other. The total number of clumps in the disk that would eventually pass through our LOS can be estimated as:
\begin{equation}
N_{\rm los} \approx \frac{P}{80~{\rm days}} N_{\rm obs} \simeq 4.5 \, a^{3/2} \, M_{\ast}^{-1/2} \, N_{\rm obs},
\label{eqn-los}
\end{equation}
where $N_{\rm obs}$ is the number of dips seen along our LOS during the 80-day K2/C2 observing period. To get the total number of clumps in the disk, $N_{los}$ must also be corrected for the fact that the LOS only intercepts $R_{\ast}/H$ of the disk for edge-on inclinations, where $H$ is the scale height of the clump distribution in solar radii. Taking the disk scale height as $H/a\sim0.1$, and using $N_{\rm obs}\sim15$  (from $N_{\rm dip}$ in Table~\ref{tab-properties}), the total number of clumps in the disk is:
\begin{equation}
N_{\rm tot} \approx 1000 \, a^{5/2} \, M_{\ast}^{-1/2} \, R_{\ast}^{-1}.
\label{eqn-total}
\end{equation}
This predicts that $\sim$16,000 clumps exist at $\sim$2 AU for a star with $M_{\ast} = 0.5~M_\odot$ and $R_{\ast}=0.5~R_\odot$. This corresponds to $\sim$4 $M_{\oplus}$ or $\sim$20\% of the minimum mass solar nebula at that disk radius, which is fairly reasonable when compared to planetesimal formation models that predict $\sim$10\% of disk mass is converted into planetesimals \cite[e.g.,][]{2011A&A...529A..62J}. Moreover it is implicit in many terrestrial planet formation models that much of the solid mass in protoplanetary disks is in planetesimals that eventually coalesces to form terrestrial planets.

Another concern is that the streaming instability operates in regions with high dust-to-gas ratios, namely the disk mid-plane where settling increases the dust density. This conflicts with the low extinction toward the dippers (Table~\ref{tab-properties}), which implies that we are not seeing their disks completely edge-on, and thus the transiting clumps are located above the disk mid-plane. However, high dust-to-gas ratios can also arise from gas depletion, thus the necessary environment for streaming instability may be possible at high altitudes if the gas has been sufficiently dispersed. Note that low disk densities are not necessarily an impediment to this process; for example, \cite{2015A&A...579A..43C} found that the region in which planetesimal formation occurs should just move closer to the star as the disk dissipates. It could also be that planetesimals are scattered to high altitudes due to interactions with nearby planets \cite[e.g.,][]{2011A&A...531A..80K} or that the disk is warped due to an embedded inclined protoplanet \cite[e.g.,][]{2014MNRAS.442.3700F}.

\section{Summary \& Future Work \label{sec-summary}}

We have presented a sample of ten young ($\lesssim$10 Myr) late-K and M dwarf ``dipper" stars located in the Upper Sco and $\rho$ Oph star-forming regions. These sources were identified by dimming events in their K2/C2 light curves, which exhibited $\sim$10--20 dips in flux over the 80-day observing campaign with typical durations of $\sim$0.5--2 days and depths of up to $\sim$40\%. We classified these dippers as either quasi-periodic or aperiodic: quasi-periodic refers to sources whose dips occur at periodic intervals but with varying shapes and depths; aperiodic refers to sources whose dips appear stochastically and with varying shapes and depths. 

Our multi-wavelength follow-up observations revealed Li {\sc I} absorption and H$\alpha$ emission consistent with stellar youth, but also rates of accretion spanning those expected from classical and weak-line T Tauri stars. All the dippers in our sample showed IR excesses consistent with protoplanetary disks, although our sub-mm observations implied low disk masses between those of typical protoplanetary and debris disks. SED modeling suggested that most dippers in our sample have inner disks extending to within a few stellar radii, although we could not rule out gaps in the disks, and two sources likely host disks with inner holes. 

We found a positive correlation between dip depth and WISE-2 excess (but {\em not} WISE-3/4 excesses). We interpreted this as evidence that the dipper phenomenon arises from the inner disk near the co-rotation radius, where dust temperatures reach $\sim$600 K. However, this was difficult to reconcile with the aperiodic dippers in our sample because disk structures so close to the star should result in quasi-periodic dimming events. Although previous studies have reconciled aperiodic dippers among CTTS populations with unstable accretion regimes, the dippers presented in this work were mostly WTTS with low disk masses and weak accretion signatures.

We therefore explored several mechanisms that could be driving the dipper phenomenon: (1) inner disk warps near the co-rotation radius related to accretion; (2) RWI-driven vortices at the inner disk edge; and (3) transiting clumps of circumstellar material related to planetesimal formation at a few AU. For the quasi-periodic dippers, the transiting clump scenario is clearly ruled out as orbits near the co-rotation radius are needed to reproduce the regular dimming events with short periods. Rather, the quasi-periodic dippers are likely explained by occulting inner disk warps or RWI-driven vortices; however, the requirement of accretion in the disk warp scenario may make RWI-driven vortices a more likely explanation for the quasi-periodic dippers in our sample with low disk masses and weak accretion signatures. Given the diversity of the dipper light curves, it could be that the dippers are produced by different mechanisms that correspond to different stages of disk evolution. 

Additional follow-up observations may help distinguish among the proposed mechanisms driving the dipper phenomenon. High-resolution sub-mm imagery will reveal disk structures and geometries, while high-resolution spectra will better constrain stellar rotation periods and accretion states. Multi-wavelength, time-series photometry will also help determine dust grain properties. Moreover, follow-up observations of the 15 dippers found in subsequent visual inspection of the K2/C2 light curves will more than double our sample size. Such detailed analyses of dipper populations in young star-forming regions are important as they provide a unique opportunity to study inner disk structure and dynamics during key epochs of planet formation.

\begin{acknowledgements}
MCA and JPW were supported by NSF and NASA grants AST-1208911 and NNX15AC92G, respectively. EG was supported by an International Short Visit award from the Swiss National Science Foundation and by the Institute for Theory and Computation at Harvard CfA. MCW and GMK acknowledge European Union support through ERC grant 279973. TSB was supported through NASA grants ADAP12-0172 and ADAP14-0245. DML and TLJ thank Allan R. Schmitt for providing the light curve analysis software LcTools used in this study. We thank Fei Dai for checking light curves for planetary signals and Roberto Sanchis-Ojeda for providing us with photometry from his pipeline for many of our targets. This work was based on observations obtained with the Infrared Telescope Facility, which is operated by the University of Hawaii under Cooperative Agreement NNX-08AE38A with the NASA Science Mission Directorate's Planetary Astronomy Program.
\end{acknowledgements}

\bibliography{ms.bib}

\clearpage
\begin{turnpage}
\capstartfalse                                  
\begin{deluxetable*}{rrrrrrrrrrrrrr}                                 
\tabletypesize{\scriptsize}                                           
\tablewidth{0pt}                                                         
\tablecaption{Photometry of K2/C2 Dippers \label{tab-phot}}                  
\tablecolumns{14}                                                         
\tablehead{                                       
   \colhead{EPIC ID}                                      
 & \colhead{R$_{\rm c}$\textsuperscript{a}}                                                          
 & \colhead{I$_{\rm c}$\textsuperscript{a}}                                                          
 & \colhead{2MASS-J}                                                          
 & \colhead{2MASS-H}                                                          
 & \colhead{2MASS-K$_{\rm S}$}                                                          
 & \colhead{WISE-1}                                                         
 & \colhead{WISE-2}                                                   
 & \colhead{WISE-3}                                                   
 & \colhead{WISE-4}                                                                                               
 & \colhead{PACS-70}                                                                                               
 & \colhead{PACS-100}                                                                                               
 & \colhead{PACS-160}                                                                                               
 & \colhead{SMA\textsuperscript{b}} \\
   \colhead{} 
 & \colhead{(mag)}                                     
 & \colhead{(mag)}                                     
 & \colhead{(mag)}                                     
 & \colhead{(mag)}                                     
 & \colhead{(mag)}                                     
 & \colhead{(mag)}                                     
 & \colhead{(mag)}                                     
 & \colhead{(mag)}                                     
 & \colhead{(mag)}                                     
 & \colhead{(mJy)}
 & \colhead{(mJy)}
 & \colhead{(mJy)}
 & \colhead{(mJy)}                                  
}                               
\startdata                                                               
203343161 & 15.95$\pm$0.01 & 14.06$\pm$0.01 & 12.17$\pm$0.02 & 11.53$\pm$0.02 & 11.17$\pm$0.02 & 10.76$\pm$0.02 & 10.27$\pm$0.02 & 8.78$\pm$0.03 & 7.24$\pm$0.13 & --- & --- & --- & $<5.7$  \\   
203410665 & 10.50$\pm$0.01 &   9.71$\pm$0.01 &   8.69$\pm$0.02 &   7.95$\pm$0.06 &   7.52$\pm$0.02 &   6.81$\pm$0.07 &   6.49$\pm$0.02 & 4.84$\pm$0.02 & 2.67$\pm$0.02 & --- & --- & --- & $<4.8$  \\            
203895983 & 12.49$\pm$0.01 & 11.19$\pm$0.01 &   9.98$\pm$0.03 &   9.22$\pm$0.03 &   8.85$\pm$0.02 &   8.61$\pm$0.02 &   8.23$\pm$0.02 & 6.57$\pm$0.02 & 4.52$\pm$0.03 & --- & --- &  --- & $<4.5$  \\            
203937317 & 12.76$\pm$0.01 & 11.54$\pm$0.01 &   9.65$\pm$0.03 &   8.61$\pm$0.04 &   8.06$\pm$0.02 &   7.55$\pm$0.03 &   7.17$\pm$0.02 & 5.15$\pm$0.02 & 3.13$\pm$0.04 & 310$\pm$15 & --- & --- & 10.3$\pm$1.8 \\            
204137184 & 15.11$\pm$0.01 & 13.33$\pm$0.01 & 11.73$\pm$0.02 & 11.04$\pm$0.02 & 10.67$\pm$0.02 & 10.08$\pm$0.03 &   9.43$\pm$0.02 & 7.96$\pm$0.02 & 5.96$\pm$0.05 & --- & --- & --- & $<4.5$  \\            
204630363 & 12.11$\pm$0.01 & 11.26$\pm$0.01 & 10.07$\pm$0.03 &   9.33$\pm$0.03 &   8.95$\pm$0.03 &   8.47$\pm$0.02 &   8.23$\pm$0.02 & 6.25$\pm$0.02 & 3.64$\pm$0.02 & --- & --- & --- & 28.0$\pm$1.5 \\            
204757338 & 15.12$\pm$0.02 & 13.17$\pm$0.02 & 11.23$\pm$0.02 & 10.57$\pm$0.02 & 10.22$\pm$0.02 & 10.02$\pm$0.02 &   9.73$\pm$0.02 & 8.90$\pm$0.03 & 6.59$\pm$0.10 & --- & --- & --- & 12.5$\pm$1.4 \\           
204932990 & 14.71$\pm$0.01 & 13.10$\pm$0.01 & 11.45$\pm$0.02 & 10.76$\pm$0.03 &  10.40$\pm$0.02 & 10.08$\pm$0.02 &  9.70$\pm$0.02 & 8.17$\pm$0.03 & 5.95$\pm$0.05 & --- & --- & --- & $<4.8$   \\            
205151387 & 12.61$\pm$0.01 & 11.51$\pm$0.01 & 10.22$\pm$0.02 &   9.48$\pm$0.02 &    9.15$\pm$0.03 &  8.70$\pm$0.02  &  8.44$\pm$0.02 & 6.17$\pm$0.02 & 3.62$\pm$0.02 & 306$\pm$7 & 356$\pm$6 & 397$\pm$25 & 12.3$\pm$2.0 \\              
205519771 & 14.79$\pm$0.02 & 13.19$\pm$0.02 & 11.75$\pm$0.03 & 11.05$\pm$0.05 &  10.75$\pm$0.04 & 10.56$\pm$0.02 & 10.25$\pm$0.02 & 8.10$\pm$0.02 & 6.47$\pm$0.07 &   ---   &   ---  &   ---   & $<3.9$            
\enddata    
\tablenotetext{a}{Cousins magnitudes estimated from our SNIFS optical spectra (Section~\ref{sec-spectra}) using the revised filter profiles and zero points from \cite{2015PASP..127..102M}.}
\tablenotetext{a}{Detected fluxes or 3$\sigma$ upper limits from our SMA 1.3 mm observations (Section~\ref{sec-sma}), except for EPIC 205151387 whose SMA 1.3mm measurement comes from \cite{2008ApJ...686L.115C} and was also detected by ALMA at 880 $\mu$m with 47.28$\pm$0.91 mJy \citep{2014ApJ...787...42C}.}
\end{deluxetable*}                                                        
\capstartfalse
\end{turnpage}
\clearpage
\global\pdfpageattr\expandafter{\the\pdfpageattr/Rotate 90}

\end{document}